\documentclass[prd,a4paper,twocolumn,superscriptaddress,floatfix,longbibliography]{revtex4-2}

\usepackage{amsfonts}
\usepackage{amsmath}
\usepackage{graphicx}
\usepackage{amssymb}
\usepackage{color}
\usepackage{subfigure}
\usepackage{mathtools}
\usepackage{tikz-cd}
\usepackage[colorlinks]{hyperref}
\usepackage[normalem]{ulem}
\usepackage{bbm}

\begin{document}
\title{Closed-time-path approach to the optomechanical back-reaction problem}

\author{Salvatore Butera}
\affiliation{School of Physics and Astronomy, University of Glasgow, Glasgow G12 8QQ, United Kingdom}

\begin{abstract}
We present a perturbative closed-time-path (in-in) formulation of an optomechanical system in which a quantum field interacts with a moving mirror via radiation pressure. We derive the effective action governing the dynamics of the moving mirror, incorporating the full back-reaction of the cavity field. These effects are encoded in fluctuation and dissipation kernels, that we show satisfy fluctuation-dissipation relations, and whose spectral structure reveals a direct connection with the underlying physical mechanism responsible for the back-reaction, that is particle creation by the dynamical Casimir effect. By deriving the semiclassical equations of motion for the moving mirror, and computing the energy radiated into the field within the in-out formalism of quantum field theory, we verify the energy balance between the mechanical energy dissipated by the optical back-reaction forces acting on the mirror and the energy carried by the particles created in the field.
\end{abstract}
\maketitle

\section{Introduction}

The study of quantum fields interacting with movable boundaries has attracted sustained interest within the scientific community since the pioneering works by Moore~\cite{Moore-DCE-1970}, Fulling and Davies~\cite{Fulling_Davies-DCE,Davies-Fulling-BH} in the 1970's. These authors demonstrated that vacuum fluctuations of quantum fields can be parametrically amplified by mechanical motion, resulting in the spontaneous creation of real particles from the vacuum. This phenomenon, know as dynamical Casimir effect (DCE)~\cite{Yablonovitch-DCE,Schwinger-DCE} has since been investigated in a wide range of analogue settings, including condensed-matter, atomic and optical system  (see, e.g., the topical reviews~\cite{Dodonov-DCE-2001,Dodonov-DCE-2010,Dodonov-DCE-2020}), and found experimental confirmation in superconducting electronic circuits~\cite{Wilson-DCE-Analog-2011,Lahteenmaki-DCE-Analog-2013,Nation-RMP-2012}.
While much of the work on the DCE has been carried out at the test-field level, namely assuming the motion of the boundary to be externally prescribed, the theoretical investigation of \emph{back-reaction} effects, where the vacuum itself influences the motion of mechanical objects, has received comparatively less attention, mainly due to the significant challenges this problem poses.
Early studies in this direction, incorporating the \emph{dynamics} of the moving boundary and its interaction with quantum fields, revealed that vacuum fluctuations can influence the motion of mechanical objects~\cite{KardarRMP1999}, inducing not only dissipation and fluctuations~\cite{Reynaud-QuantOpt-1992,Reynaud-JdP-1993,Reynaud-PhysLettA-1993,Unruh-PRD-2014,Savasta-PRX-2018,Butera-PRA-2019} but also quantum decoherence~\cite{Dalvit-PRL-2000,MaiaNeto-PRA-2000,Butera-EPL-2019,Butera2023}. These efforts built upon pioneering work on the back-reaction problem in the context of early universe cosmology, which demonstrated that quantum matter fields back-act on and affect the spacetime dynamics, as a result of cosmological particle creation~\cite{Hu-BR-Cosm-1978, Hu-BR-Cosm-1979-I, Hu-BR-Cosm-1979-II, Hu-BR-Cosm-1979-III, Hartle-BR-Cosm-1980-IV, Hartle-BR-Cosm-1981-V,Hu-BR-InIn-1987, Campos-BR-InIn-1990, Hu-BR-InIn-1994, Campos-BR-InIn-1994}.

While the investigation of quantum fields in the presence of movable boundaries was originally motivated by fundamental questions about the physics of the quantum vacuum and its instability in nonequilibrium configurations~\cite{milonni-book}, in recent decades this topic has acquired increasing technological relevance, particularly through its direct connections to quantum optomechanics~\cite{Aspelmeyer_RMP}. This field, which explores both the fundamental principles and practical applications of the interplay between light and mechanical motion, has expanded rapidly over the past few decades, initially driven by the interest in probing genuine macroscopic quantum behaviour~\cite{Leggett2002,Marshall2003} and its applications in the physics of gravitational-wave detectors~\cite{Abramovici-science-1992, TAccadia_2012, Danilishin2012}. Recent experimental advances, that have enabled the cooling of macroscopic oscillators to the quantum level~\cite{Teufel2011,OConnell-Science-2010,chan2011laser,Aspelmeyer-Science-2020}, have dramatically broadened the reach of optomechanical devices, which are now regarded as one of the most promising platforms for the development of future quantum technologies~\cite{Paternostro_Review}.
Their use has been explored, for instance, in quantum metrology and sensing~\cite{stange2021science}, in quantum thermodynamics, as a platform to implement quantum thermal machines~\cite{Nori-QuantumHEatEng-2007,Bariani-QuantumHEatEng-I,Bariani-QuantumHEatEng-II,Bariani-QuantumHEatEng-III,Nori-QuantumHEatEng-2023} and, due to the exceptionally long coherence times of acoustic resonators~\cite{MacCabe2020,Beccari2022}, in quantum information storage~\cite{MechQbit-PRX-2021,navarathna2022good} and processing~\cite{Rabl2012}.

These considerations highlight both the fundamental and technological relevance of the physics of the quantum vacuum, and in particular of the problem of the back-reaction of quantum fields on macroscopic objects. From a fundamental perspective, incorporating these effects is essential for developing fully self-consistent and energy-conserving theories. More broadly, clarifying the nonlinear processes through which macroscopic quantum behaviour emerges from microscopic fluctuations is poised to deliver deeper insight into the principles governing the boundary between the classical and quantum domains~\cite{Zurek1991}. From a more practical perspective, an accurate understanding of the processes driving the mutual interaction between the optical and mechanical degrees-of-freedom is pivotal to achieve full quantum control of optomechanical devices and exploit these systems in concrete technological applications.

Motivated by this dual perspective, the objective of this paper is to lay the foundational work for a general and fully self-consistent description of the back-reaction of quantum fields on moving objects, such as optical mirrors. The formalism developed here, and extended in planned follow-up works, aims to provide a comprehensive understanding of back-reaction physics starting from first principles, and deliver refined and powerful mathematical tools to effectively describe the nonlinear dynamics of optomechanical systems and their mathematically equivalent circuit-QED configurations~\cite{Johansson-PRL-2009,Johansson-PRA-2010}. We perform this investigation describing the coupling between a radiation field and a moving mirror through the theory of radiation pressure interaction~\cite{Law-MirFieldInt-1995,Law-MirFieldInt-2011,Butera-2025-corrections}, widely adopted within the scientific community. Building upon previous efforts~\cite{Butera-IF,Butera2023}, we pursue an open quantum system description based on the Feynman--Vernon influence functional formalism~\cite{Feynman-IF}, a functional method based on the Keldysh--Schwinger closed-time-path formalism~\cite{Schwinger1961,Keldysh1965} that has proved effective in similar studies of back-reaction in cosmological settings~\cite{Hu-BR-InIn-1987, Campos-BR-InIn-1990, Hu-BR-InIn-1994, Campos-BR-InIn-1994}.

The results presented in this paper extend previous work by the author~\cite{Butera-IF,Butera2023}, by providing a more detailed and comprehensive analysis of the fluctuation and dissipation kernels that encode the back-reaction of the quantum field on the moving mirror. Specifically, we carry out a spectral analysis of the effective action for the mirror, showing that its structure allows for a direct interpretation of the back-reaction in terms of the underlying DCE mechanism. A central result of this paper is the demonstration of the balance between the energy dissipated by the optical back-reaction forces acting on the moving mirror and the energy gained by the field through spontaneous particle creation processes, thus confirming the self-consistency of the theory. To avoid unnecessary technical complications while preserving conceptual clarity, we adopt the working assumptions of~\cite{Law-MirFieldInt-1995,Law-MirFieldInt-2011} and consider a one-dimensional optical cavity with a single moving mirror. Extending these results to more experimentally realistic three-dimensional configurations will be the subject of future work, where we will also derive the nonlinear Langevin equation describing the quantum Brownian motion of the mirror coupled to the cavity field.

The paper is structured as follows: In Sec.~\ref{Sec:radiation_pressure}, we review the theory of radiation pressure used to describe the mutual interaction between the radiation field and the moving mirror, and establish the notation adopted throughout the paper. In Sec.~\ref{sec:IF-IA}, we introduce the theory of Feynman--Vernon influence functional and the corresponding influence action, which constitutes the open quantum system framework employed to describe the effects of the field on the mirror dynamics. In doing so, we also present the basic elements of the Keldysh--Schwinger closed-time-path (CTP) formalism this method builds upon. Because of the nonlinear structure of the radiation pressure interaction, our analysis is carried out pursuing a perturbative approach, which we outline in the same section. To make this paper as self-contained as possible we present, in Sec.~\ref{Sec:Wick}, a brief digression on Wick's theorem for Gaussian states, of which we make extensive use in the following sections to evaluate the field correlations that contribute to the mirror influence action. This action is calculated and discussed in depth in Sec.~\ref{sec:rad_pres_infl_func}. There, we show that the first-order contribution in the perturbative expansion gives rise to the static Casimir potential~\cite{Butera-IF}, while the second-order contribution encodes the dynamical back-reaction of the quantum field on the mirror. The latter takes the form of fluctuation and dissipation effects, accounted for by corresponding kernel functions in the final expression of the effective action. We perform a spectral analysis of the effective action, demonstrating the connection between the field back-reaction and the emission of photon pairs by DCE. In Sec.~\ref{sec:eom}, we derive the semiclassical equation of motion for the mirror displacement, identifying the optical back-reaction forces that we then use to calculate the explicit expression for the dissipated mechanical energy. Once again, the spectral structure of the dissipated energy is shown to carry the hallmark of DCE emission. We demonstrate the energy balance between the energy lost by the mirror and the energy gained by the field through spontaneous particle creation in Sec.~\ref{sec:balance}, where the latter is calculated by evaluating the persistence amplitude of the initial field state within the standard in-out formalism of quantum field theory. The demonstration of energy balance confirms the full self-consistency of the developed formalism. We draw our conclusions in Sec.~\ref{sec:conclusions}.

\section{Radiation pressure interaction\label{Sec:radiation_pressure}}

\begin{figure}[t]
    \centering
    \includegraphics[width = 3.5 in]{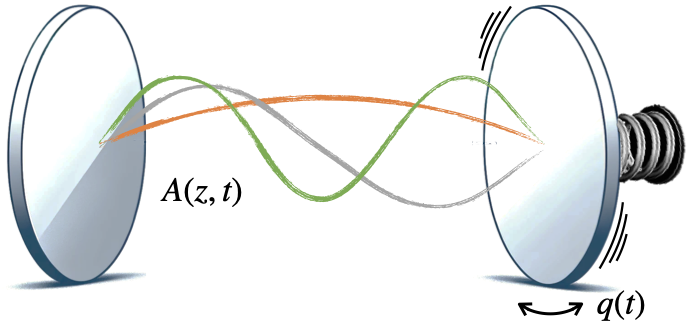}
    \caption{Schematic representation of the optical cavity: the left mirror is fixed at the position $z=0$, while the right mirror, whose time-dependent position is $z=q(t)$, is free to move within a the potential $V(q)$. The mirrors interact with the scalar field $A(z,t)$ enclosed in the cavity.}
\label{Fig1}
\end{figure}

We pursue our study by considering the system sketched in Fig.~\ref{Fig1}, that is a one-dimensional optical cavity bounded by two mirrors enclosing the scalar field $A(z,t)$. In such a one-dimensional configuration, the scalar field plays the role of the electromagnetic vector potential. We assume the left mirror fixed at the position $z=0$, while the right mirror has mass $m$ and moves within the confining potential $V(q)$. We indicate the time-dependent position of the moving mirror as $q(t)$.
For the sake of simplicity, we work within the same approximation as in~\cite{Law-MirFieldInt-1995, Law-MirFieldInt-2011}, namely, we neglect the field outside the cavity. Strictly speaking, this assumption is physically justified provided there is an appreciable photon population within the cavity, so that the internal field dominates over the external one. Extending the model to include the degrees-of-freedom of the outside field is straightforward and does not add any conceptual novelty beyond the results presented in this paper.

The mirrors are coupled to the cavity field through the radiation pressure interaction. Following Law’s model~\cite{Law-MirFieldInt-1995, Law-MirFieldInt-2011}, this coupling is implemented by imposing vanishing (Dirichlet) boundary conditions on the field at the mirror positions
\begin{equation}
    A(z=0,t) \; = \; A(z=q(t),t) \;=\; 0,
\end{equation}
which corresponds to the idealized limit of perfectly reflecting mirrors with divergent electric susceptibility~\cite{MOF1, MOF2}. This assumption is physically justified for optical modes with frequencies below the plasma frequency $\omega_{\rm{pl}}$ of the mirror material, which therefore sets a natural high-frequency cut-off for the modes participating in the interaction.

The action $S[q,A]$ governing the dynamics of the mirror-field system is composed of the action $S_{\mathrm{M}}[q]$ for the free mirror within its confining potential and the action $S_{\mathrm{A}}'[q,A]$ that accounts for the dynamics of the cavity field:
\begin{equation}
    S[q,A] = S_{\mathrm{M}}[q] + S_{\mathrm{A}}'[q,A] .
\end{equation}
Since the cavity length equals the mirror coordinate $q(t)$, the field action acquires an explicit dependence on the mirror position. Modeling the latter as a nonrelativistic massive particle, and working in natural units $(c =\hbar = 1)$, these actions take the form
\begin{align}
    S_{\mathrm{M}}[q] &= \int_{t_i}^{t_f} dt\,\left[\frac{1}{2}m\dot{q}^2 - V(q)\right], \label{eq:SM} \\[2pt]
    S_{\mathrm{A}}'[q,A] &= \frac{1}{2}\int_{t_i}^{t_f} dt \int_{0}^{q(t)} dz\,\left[\left(\partial_t A\right)^2 - \left(\partial_z A\right)^2\right], \label{eq:SA'}
\end{align}
where overdots denote total time derivatives, $\dot{q} \equiv dq/dt$, and $t_i$ and $t_f$ are the initial and final times of the motion, respectively. Throughout this work, the explicit space and time dependence of physical quantities is suppressed in equations for readability, whenever no ambiguity arises.

Proceeding as in ~\cite{Law-MirFieldInt-1995, Law-MirFieldInt-2011}, we now expand the field in terms of the complete set of cavity eigenmodes $\varphi_k(z,t)$, with $k \in \mathbb{N}$, that instantaneously verify the boundary conditions:
\begin{equation}
    A(z,t) \; = \; \sum_{k=1}^{+\infty} Q_k(t) \,\varphi_k(z,t).\label{eq:A}
\end{equation}
Here, $Q_k(t)$ are the set of time-dependent mode amplitudes, while the field modes take the explicit form:
\begin{equation}
    \varphi_k(z,t) \equiv \sqrt{\frac{2}{q(t)}}\,\sin\!\left(\omega_k[q(t)]\,z\right),\label{eq:modes}
\end{equation}
with $\omega_k[q(t)] \equiv {k \pi}/{q(t)}$ the cavity-length-dependent modes frequencies. Note that the implicit time-dependence of the field eigenmodes is inherited from their dependence on the mirror position $q(t)$.

Substituting Eqs.~\eqref{eq:A} and \eqref{eq:modes} in Eq.~\eqref{eq:SA'}, and performing the spatial integrals, the field action $S_{\mathrm{A}}'$ can be expressed in terms of the complete set of modes amplitudes as:
\begin{multline}
    S_{\mathrm{A}}'[q,\{Q_k\}] =  \sum_{k=1}^{+\infty}\;\int_{t_i}^{t_f}\!dt\,\bigg\{\frac{1}{2} \big( \dot{Q}_k^2 - \omega_k^2(q) \,Q_k^2 \big) \\
    +\, \frac{\dot{q}}{q} \dot{Q}_k \, Q_k^{(1)}
- \frac{\dot{q}^2}{2 q^2}  Q_k \, Q_k^{(2)}\bigg\}.\label{SQ'}
\end{multline}
Here, $\{Q_k\}$ is a shorthand to denote the collection of all mode amplitudes, while $Q_k^{(1)}$, $Q_k^{(2)}$ are defined according to the recursive relation~\cite{Butera-2025-corrections}:
\begin{subequations}
\begin{align}
	Q_k^{(0)} &\equiv Q_k, \label{Q0}\\
	Q_k^{(n+1)} &\equiv \sum_{j=1}^{+\infty} g_{jk} Q_j^{(n)},\label{Qn}
\end{align}
\end{subequations}
for $n =0,1$, with the coupling coefficients:
\begin{align}
	g_{jk} &\equiv q \int_0^q dz\left(\varphi_k\frac{\partial\varphi_j}{\partial q}\right) \nonumber\\
	& = \left\{
        \begin{array}{cc}
                (-1)^{k+j}\left(\frac{2 k j}{ k^2- j^2}\right) & \quad (k\neq j) \\
                0 & \quad (k = j)
    	\end{array}.
    \right.\label{g_kj}
\end{align}
These coefficients quantify the overlap between distinct cavity modes induced by the mirror displacement. As discussed in~\cite{Butera-2025-corrections}, $Q_k^{(n)}$ can be interpreted as the fluctuation amplitude of the $k$-th mode arising from its hybridization with the other field modes, driven by mechanical fluctuations of order $n$. We remark that, to obtain Eq.~\eqref{SQ'}, we also used the following relation, that results from the completeness of the instantaneous basis:
\begin{equation}
\sum_{s=1}^{+\infty} g_{ks}g_{js} = q^2 \int_0^{q} dz \, \frac{\partial\varphi_k(z)}{\partial q} \frac{\partial\varphi_j(z)}{\partial q}.\label{compl}
\end{equation}

In preparation for the perturbative treatment of the radiation pressure interaction carried out in the following sections, it is convenient to rewrite the total action $S[q,\{Q_k\}]$ so as to explicitly separate the free dynamics of the mirror and the field, the latter defined with respect to the equilibrium cavity length within the potential $V(q)$, from their interaction. This task is achieved by decomposing the field action as:
\begin{equation}
	S_{\mathrm{A}}'[q,\{Q_k\}] = S_{\mathrm{A}}[\{Q_k\}] + S_{\rm int}[q,\{Q_k\}],
	\label{SQ'_dec}
\end{equation}
where
\begin{equation}
    S_{\mathrm{A}}[\{Q_k\}] = \frac{1}{2} \sum_{k=1}^{+\infty}\,\int_{t_i}^{t_f}{dt\,{\big(\dot{Q}_k^2 - \omega_{k0}^2Q_k^2\big)}},\label{SA0}
\end{equation}
is the action of the free field evaluated at the equilibrium length $d$ of the cavity, with $\omega_{k0} \equiv k\pi/d$ the frequency of the $k$-th mode at equilibrium, while
\begin{multline}
  S_{\rm int}[\{Q_k\},q] = \sum_{k=1}^{+\infty}\,\int_{t_i}^{t_f}dt\,\bigg\{\frac{1}{2} \big[ \omega_{k0}^2 - \omega_k^2(q) \big]\,Q_k^2 \\
    \frac{\dot{q}}{q} \dot{Q}_k \, Q_k^{(1)}
- \frac{\dot{q}^2}{2 q^2}  Q_k \, Q_k^{(2)}
\bigg\}\label{Sint}
\end{multline}
accounts for the interaction between the optical and mechanical degrees-of-freedom. Having isolated the interaction, the action for the mirror-field system is finally written in the form:
\begin{equation}
    S[q,\{Q_k\}] = S_{\mathrm{M}}[q] + S_{\mathrm{A}}[\{Q_k\}]+ S_{\rm int}[q,\{Q_k\}].\label{SqQ}
\end{equation}
For later convenience, we further decompose the interaction action in such a way to identify terms of different order in the mirror's velocity $\dot q$:
\begin{multline}
    S_{\rm int}[q,\{Q_k\}] \equiv S_{\rm int}^{(0)}[q,\{Q_k\}] + S_{\rm int}^{(1)}[q,\{Q_k\}] \\+ S_{\rm int}^{(2)}[q,\{Q_k\}], \label{Sint_012}
\end{multline}
with
\begin{subequations}
\begin{align}
    S_{\rm int}^{(0)}[q,\{Q_k\}] &= \frac{1}{2} \sum_{k=1}^{+\infty} \big[ \omega_{k0}^2 - \omega_k^2(q) \big]\,Q_k^2 \label{Sint_0},\\
    S_{\rm int}^{(1)}[q,\{Q_k\}] &= \frac{\dot{q}}{q} \sum_{k=1}^{+\infty} \dot{Q}_k \, Q_k^{(1)},\label{Sint_1}\\
    S_{\rm int}^{(2)}[q,\{Q_k\}] &= - \frac{\dot{q}^2}{2 q^2} \sum_{k=1}^{+\infty} Q_k \, Q_k^{(2)}, \label{Sint_2}
\end{align}
\end{subequations}
the components of zeroth, first, and second order in $\dot{q}$, respectively. The zeroth-order term $S_{\mathrm{int}}^{(0)}$ is independent of the mirror velocity and encodes the purely position-dependent part of the interaction. It arises from the mismatch between the equilibrium mode frequencies $\omega_{k0}$ and the instantaneous mode frequencies $\omega_k[q(t)]$. The remaining terms $S_{\mathrm{int}}^{(1)}$ and $S_{\mathrm{int}}^{(2)}$, which depend explicitly on the mirror velocity, account for strictly non-adiabatic dynamical effects of the interaction.

\section{Theory of the influence functional and action\label{sec:IF-IA}}
The objective of this paper is to provide a theoretical description of the dynamical evolution of the quantum state of the moving mirror coupled to the cavity field through the radiation pressure interaction introduced in the previous section. As discussed in the Introduction, we pursue this goal adopting an approach based on the theory of Feynman–Vernon influence functionals~\cite{Feynman-IF,Feynman-book-PathInt,Calzetta_PhysA}. This provides an open quantum systems framework for describing the nonequilibrium dynamics of interacting multi-mode systems, based on the Keldysh–Schwinger formalism~\cite{Schwinger1961,Keldysh1965,Chou1985} of CTP integrals. In what follows, we briefly review the essential elements of this formalism, starting in the next section by introducing the reduced density operator of the moving mirror.

\subsection{Reduced density operator for the moving mirror}
The quantum state of the mirror is formally described by its reduced density operator, that we indicate as $\hat\rho_{\mathrm{M}}$. This is obtained by performing the partial trace with respect to the degrees-of-freedom of the field on the density operator $\hat{\rho}$ of the combined mirror-field system:
\begin{equation}
  \hat{\rho}_{\mathrm{M}}(t) \equiv {\rm Tr}_{\mathrm{A}} [\hat{\rho}(t)].    
\end{equation}
Indicating with $\hat{U}(t_f,t_i)$ the operator that evolves the state of the combined system from the initial time $t_i$ to the final time $t_f$, then the evolved state of the mirror reads:
\begin{equation}
    \hat\rho_{\mathrm{M}}(t_f) = \text{Tr}_{\mathrm{A}} \!\left( \hat U(t_f, t_i) \, \hat \rho(t_i) \, \hat U^{\dagger}(t_f, t_i) \right).\label{rho_M_red}
\end{equation}
For the sake of the following discussion, we introduce the basis composed by the states $| q,\{Q_k\}\rangle \equiv | q\rangle \otimes |\{Q_k\}\rangle $ which are simultaneous eigenstates of the mirror position operator and the field mode amplitudes operators, with eigenvalues $q$ and $\{Q_k\}$, respectively. In this basis, the resolution of identity for the global Hilbert space reads:
\begin{equation}
    \int_{-\infty}^{+\infty} dq \, \int_{-\infty}^{+\infty}d\{Q_k\} \, | q,\{Q_k\}\rangle \langle q, \{Q_k\}| = \mathbbm{1}_q \otimes \mathbbm{1}_{\mathrm{A}}, \label{res_id}
\end{equation}
where we used the shorthand
\begin{equation}
    \int_{-\infty}^{+\infty}d\{Q_k\}\equiv\prod_{k=1}^{+\infty} \int_{-\infty}^{+\infty}dQ_k \label{int_shorthand}
\end{equation}
and introduced the identity operators $\mathbbm{1}_q$ and $\mathbbm{1}_{\mathrm{A}}$ relative to the mirror and field sub-spaces, respectively. By inserting the resolution of identity twice in Eq.~\eqref{rho_M_red}, the projection onto the position-amplitude basis yields
\begin{widetext}
\begin{align}
\rho_{\mathrm{M}}(q_f, q_f'; t_f) &\equiv \langle q_f|\hat{\rho}_{\mathrm{M}}(t)| q_f'\rangle \nonumber\\[2mm]
&= \int d\{Q_{k f}\} \int dq_i \int dq_i' \int d\{Q_{k i}\}\int d\{Q_{k i}'\} \;
U(q_f, \{Q_{k f}\}, t_f \,|\, q_i, \{Q_{k i}\}, t_i) \nonumber\\[2mm]
&\hspace{40mm}U^*(q_f', \{Q_{k f}\}, t_f \,|\, q_i', \{Q_{k i}'\}, t_i)\;
\rho(q_i, \{Q_{k i}\};q_i', \{Q_{k i}'\}; t_i),
\label{rho_M_pos}
\end{align}
\end{widetext}
where
\begin{equation}
    \rho(q, \{Q_{k}\};q', \{Q_{k}'\}\,;\, t) \equiv \langle q,\{Q_k\}|\hat{\rho}(t)| q',\{Q_k'\}\rangle
\end{equation}
and
\begin{multline}
    U(q_f,\{Q_{kf}\}, t_f \,|\, q_i, \{Q_{ki}\}, t_i) \\[2pt]
    \equiv \langle q_f, \{Q_{kf}\}|\hat U (t_f,t_i)| q_i, \{Q_{ki}\}\rangle.
\end{multline}
are the matrix elements of the density and evolution operator of the mirror-field system, respectively. The latter represents the quantum mechanical propagator encoding the probability amplitude for the system to evolve from the configuration $(q_i,\{Q_{ki}\})$ at time $t_i$ to the configuration $(q_f,\{Q_{kf}\})$ at time $t_f$ and admits a natural representation as a path integral~\cite{Feynman-book-PathInt}:
\begin{multline}
    U(q_f, \{Q_{k f}\}, t_f \,|\, q_i, \{Q_{k i}\}, t_i)
\\ \equiv \int_{q_i,t_i}^{q_f,t_f}\mathcal{D}q \int_{\{Q_{ki}\},t_i}^{\{Q_{kf}\},t_f}\mathcal{D}\{Q_k\} \;e^{iS[q,\{Q_k\}]},\label{evol_path}
\end{multline}
where the functional integrations run over all mirror and field trajectories connecting the boundary configurations. In Eq.~\eqref{evol_path}, we used for path integrals over the field mode amplitude trajectories a similar shorthand as the one used in Eq.~\eqref{int_shorthand} for standard integrals.

\subsection{Definitions of the influence functional and influence action\label{path_int_rep}}

\begin{figure}[t!]
    \centering
    \includegraphics[width = .8\linewidth]{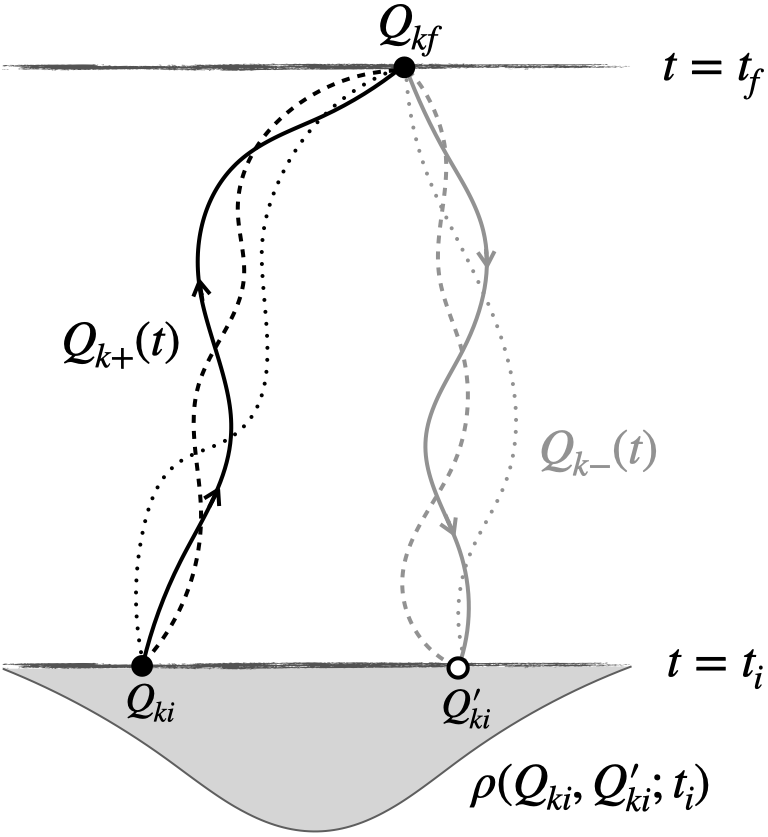}
    \caption{Illustrative representation of the closed-time trajectories that contribute to the influence functional for the moving mirror: The initial state of the field, described by the density operator $\rho_{\mathrm{A}}(Q_{ki},Q_{ki}';t_i)$, is evolved forward in time along the path $Q_{k+}(t)$, driven by the mirror motion $q_+(t)$, and backward in time along the path $Q_{k-}(t)$, driven by the motion $q_-(t)$. The forward and backward branches of the path integral meet in the same point $Q_{kf}$ at the final time $t_f$, thus closing the overall path. Integration over all possible final field amplitudes $Q_{kf}$ and against the variables $Q_{ki}$, and $Q_{ki}'$ sampling the initial state of the field, yields the influence functional, as define in Eq.~\eqref{Fqq'} in the text.}
\label{Fig:CTP_IF}
\end{figure}

Following the standard open quantum system approach, we assume that the system of interest and its environment are uncorrelated at the initial time. In the present context, these correspond to the mirror and the cavity field, respectively, so that the initial density operator of the combined system factorizes as:
\begin{equation}
\hat{\rho}(t_i) = \hat{\rho}_{\rm M}(t_i)\otimes\hat{\rho}_{\rm A}(t_i), \label{dens_op_fact}    
\end{equation}
with $\hat{\rho}_{\rm A}$ the density operator relative to the field sector. We anticipate that, in what follows, we consider the field to be initially in a thermal state at temperature $T$. By using Eqs.~\eqref{evol_path} and \eqref{dens_op_fact}, the evolved state of the mirror given in Eq.~\eqref{rho_M_pos} can be written in the form:
\begin{multline}
    \rho_{\mathrm{M}}(q_f, q_f'; t_f) = \int dq_i \int dq_i' \;\rho_{\mathrm{M}}(q_i, q_i'; t_i)\\
\times\bigg\{\int_{q_i, t_i}^{q_f, t_f} \mathcal{D}q_+\int_{q_i', t_i}^{q_f', t_f} \mathcal{D}q_- \,
e^{{i} \big(S_q[q_+] - S_q[q_-]\big)}
F[q_+,q_-]\bigg\}.\label{rho_M_IF}
\end{multline}
This equation formally introduces the \emph{influence functional} $F[q_+,q_-]$, which accounts for the effects of the cavity field on the dynamical evolution of the mirror. In operator form, this is defined as
\begin{equation}
    F[q_+,q_-] \equiv \mathrm{Tr}_{\mathrm{A}}\!\Big(
    \hat{U}_{\mathrm{A}}^{q_+}(t_f, t_i)\,\hat{\rho}_{\mathrm{A}}(t_i)\,
    \big(\hat{U}_{\mathrm{A}}^{q_-}\big)^\dagger(t_f, t_i)\Big),
    \label{eq:IF}
\end{equation}
where $\hat{U}_{\mathrm{A}}^{q}$ is the evolution operator that encodes the response of the field to the mirror motion, described by the trajectory $q(t)$. The matrix elements of $\hat{U}_{\mathrm{A}}^{q}(t_f,t_i)$ in the amplitude basis are given by the path integral
\begin{align}
U_{\mathrm{A}}^{q} &(\{Q_{k f}\}, t_f \,|\, \{Q_{k i}\}, t_i) \nonumber\\[4pt]
&=\int_{\{Q_{k i}\}, t_i}^{\{Q_{k f}\}, t_f} \mathcal{D}\{Q_k\} \;e^{{i} \big(
S_{\mathrm{A}}[\{Q_k\}] + S_{\rm int}[q, \{Q_k\}]
\big)},\label{UAq}
\end{align}
in which $q(t)$ enters as an external parameter. By using Eq.~\eqref{UAq}, the influence functional takes the explicit form:
\begin{widetext}
\begin{multline}
       F[q_+,q_-] \\[4pt]
=\int d\{Q_{k f}\}\; \int d\{Q_{k i}\} \int d\{Q_{k i}'\}\;U_{\mathrm{A}}^{q_+} (\{Q_{k f}\}, t_f \,|\, \{Q_{k i}\}, t_i) \;\; \big(U_{\mathrm{A}}^{q_-}\big)^* (\{Q_{k f}\}, t_f \,|\, \{Q_{k i}'\}, t_i)\;\;\rho_{\mathrm{A}}(Q_{k i}, Q_{k i}'; t_i). \label{Fqq'}
\end{multline}
\end{widetext}
Eq.~\eqref{Fqq'}, together with Eq.~\eqref{UAq}, allow for a natural interpretation of the influence functional as a CTP generating functional for the field, with the mirror motion acting as a time-dependent external drive coupled to the field through the interaction $S_{\mathrm{int}}[q,Q_k]$. As pictorially illustrated in Fig.~\ref{Fig:CTP_IF}, the influence functional is expressed as a sum over pairs of field trajectories defined on a closed-time contour. One branch of this contour corresponds to forward time evolution of the field along the path $Q_{k+}(t)$ starting from $Q_{ki}$ at the initial time and ending at $Q_{kf}$ at the final time $ t_f $, driven by the mirror trajectory $ q_+(t)$. The second branch, $Q_{k-}(t)$, describes instead backward time evolution of the field from the same amplitudes $Q_{kf}$ at $t_f$ to $Q_{ki}'$ at $ t_i $, driven by the mirror trajectory $q_-(t)$.  The trace over the field degrees-of-freedom enforces the closure of the contour at  $t_f$, while integration against the density matrix $\rho_{\mathrm{A}}(Q_{ki},Q_{ki}';t_i)$ samples the initial state of the field.
In contrast to the standard \emph{in–out} formalism, which is tailored to the computation of transition amplitudes between prescribed initial and final states of a physical system, the CTP approach is referred to as the \emph{in–in} formalism because integration paths are closed back at the initial time, where the average over the initial density operator is performed. At the expense of doubling the number of degrees-of-freedom, the CTP formalism allows for the direct calculation of the time evolution of expectation values of physical observables and the derivation of the corresponding real and causal equations of motion, quantities that are not readily accessible within the in--out formalism.

The influence functional encodes the full back-action of the field on the mirror dynamics. This induces nonlocal memory effects that are made explicit by introducing the \emph{influence action} $\mathcal{A}[q_+,q_-]$, defined through
\begin{equation}
    F[q_+,q_-] \equiv e^{i\mathcal{A}[q_+,q_-]},\label{IA_def}
\end{equation}
which re-expresses the effect of the field as an additional contribution to the free mirror CTP action. Using Eq.~\eqref{IA_def}, the evolved mirror density operator in Eq.~\eqref{rho_M_IF} reads
\begin{multline}
    \rho_{\mathrm{M}}(q_f, q_f'; t_f)
    =
    \int dq_i \int dq_i' \,\rho_{\mathrm{M}}(q_i, q_i'; t_i)\\
    \bigg(
    \int_{q_i, t_i}^{q_f, t_f} \mathcal{D}q_+
    \int_{q_i', t_i}^{q_f', t_f} \mathcal{D}q_- \,
    e^{i\left( S_q[q_+] - S_q[q_-] + \mathcal{A}[q_+,q_-] \right)}
    \bigg).
    \label{rho_M_IA}
\end{multline}
From this equation we notice that, while the free CTP mirror action $S_q[q_+] - S_q[q_-]$ does not mix the forward and backward trajectories, the influence action $\mathcal{A}[q_+,q_-]$ couples $q_+(t)$ and $q_-(t)$, making explicit the non-Markovian quantum character of the field-induced back-action, since the dynamics of the mirror at time $t$ depends on its entire past history through the nonlocal structure of $\mathcal{A}[q_+,q_-]$.

The discussion developed thus far is exact, as no approximations have been introduced. In general, however, an exact evaluation of the influence functional, or equivalently of the influence action, is feasible only for relatively simple configurations and is analytically intractable in closed form for more complex systems. This is the case for the optomechanical setup considered here, owing to the intrinsically nonlinear nature of the radiation pressure interaction. Therefore, we adopt in what follows a perturbative strategy, considering the interaction as a small correction to the free dynamics of the mirror and the field. With this objective in mind, we present in the next section the general perturbative construction of the influence functional and the corresponding action.

\subsection{Perturbative expansion of the influence functional and influence action\label{sec:pert_IA}}

Let us consider a generic observable ${O}(\{Q_{k+}\},\{Q_{k-}\})$ depending on both the forward- and backward-evolving field amplitudes. We define the CTP average of this observable as: 
\begin{widetext}
\begin{multline}
\big<\mathcal{O}\big(\{Q_{k+}\},\{Q_{k-}\}\big)\big>_{\rm A} \equiv \int_{-\infty}^{\infty} d\{Q_{kf}\} \int_{-\infty}^{\infty} d\{Q_{ki}\} \int_{-\infty}^{\infty} d\{Q_{ki}'\} \;\rho_{\rm A}(\{Q_{ki}\},\{Q_{ki}'\},t_i)\\
	\times\bigg\{\int_{Q_i,t_i}^{Q_f,t_f}\mathcal{D}\{Q_{k+}\}\int_{Q_i',t_i}^{Q_f',t_f}\mathcal{D}\{Q_{k-}\}\;\mathcal{O}\big(\{Q_{k+}\},\{Q_{k-}\}\big)\;e^{i\left(S_{\mathrm{A}}[\{Q_{k+}\}]-S_{\mathrm{A}}[\{Q_{k-}\}]\right)}\bigg\}.
\label{Eq:AverageDef}
\end{multline}
\end{widetext}
This definition serves us as the basis for constructing the perturbative expansion of the influence functional and the corresponding influence action. To this end, we rewrite the field propagator in Eq.~\eqref{UAq} by formally separating the free evolution from the interaction term:
\begin{align}
U_{\mathrm{A}}^{q} &(\{Q_{k f}\}, t_f \,|\, \{Q_{k i}\}, t_i) \nonumber\\[4pt]
&=\int_{\{Q_{k i}\}, t_i}^{\{Q_{k f}\}, t_f} \mathcal{D}\{Q_k\} \,e^{i S_{\rm int}[q, \{Q_k\}]}\,e^{i S_{\mathrm{A}}[\{Q_k\}]},\label{UAq2}
\end{align}
so that the interaction can be interpreted as a multiplicative weight to the free field propagator. Combining this decomposition with the definition of the influence functional in Eq.~\eqref{Fqq'}, it follows that the latter can be expressed as the CTP field average:
\begin{equation}
   F[q_+,q_-] = \big\langle e^{ i ( S_{\mathrm{int}}[q_+, Q_{k+}] - S_{\mathrm{int}}[q_-, Q_{k-}] )}\big\rangle_{\rm A}.
   \label{IF_path}
\end{equation}

This compact expression is the starting point for the perturbative expansion of the influence functional in powers of $S_{\mathrm{int}}$, which is readily obtained by power-expanding the exponential in Eq.~\eqref{IF_path}. Labelling as $F_i$ ($\mathcal{A}_i$) terms of the influence functional (action) of $i-$th order in $S_{\rm int}$, and working up to second order in the interaction we write:
\begin{align}
    F&[q_+,q_-] = 1 + F_1[q_+,q_-] + F_2[q_+,q_-] \nonumber\\[6pt]
     &= 1 + i\mathcal{A}_1[q_+,q_-] + \big(i\mathcal{A}_2[q_+,q_-] \nonumber\\[2pt]
     &\hspace{6mm}-\tfrac{1}{2}\mathcal{A}_1^2[q_+,q_-]\big)\nonumber\\[6pt]
     &=1 + i \langle S_{\rm int}[+] - S_{\rm int}[-] \rangle_{\mathrm{A}} \nonumber\\[2pt]
     &\hspace{6mm}-\tfrac{1}{2} \langle \big( S_{\rm int}[+] - S_{\rm int}[-] \big)^2 \rangle_{\mathrm{A}}.  \label{A_expanded}
\end{align}
Here, we have used the shorthand $S_{\rm int}[\pm]\equiv S_{\rm int}[q_\pm, Q_{k\pm}]$, for brevity. Identifying terms of the same order in Eq.~\eqref{A_expanded} we infer the following relations:
\begin{subequations}
\begin{align}
    F_1[q_+&,q_-] = i \mathcal{A}_1[q_+,q_-] \nonumber\\[2pt]
    &= i\,\big\langle  S_{\rm int}[+] -  S_{\rm  int}[-] \big\rangle_{\mathrm{A}},\label{F1}\\[6pt]
    F_2[q_+&,q_-] = i \mathcal{A}_2[q_+,q_-] - \tfrac{1}{2}\mathcal{A}_1^2[q_+,q_-] \nonumber\\[2pt]
    &= -\tfrac{1}{2} \big\langle \big( S_{\rm int}[+] - S_{\rm  int}[-]\big)^2 \big\rangle_{\mathrm{A}}.\label{F2}
\end{align}
\end{subequations}
These can be inverted to explicitly write each component of the effective action in terms of field correlators:
\begin{subequations}
\begin{align}
    &\mathcal{A}_{1}[q_+,q_-] = \left\langle S_{\rm int}[+] - S_{\rm int}[-] \right\rangle_{\mathrm{A}},\label{A1}\\[4pt]
    &\mathcal{A}_{2}[q_+,q_-] =\frac{i}{2}\Big[
\big\langle \big( S_{\rm int}[+] - S_{\rm int}[-] \big)^{2} \big\rangle_{\mathrm{A}}\nonumber\\[4pt]
    & \hspace{20mm}- \mathcal{A}_{1}^2[q_+,q_-]\Big]. \label{A2}
\end{align}
\end{subequations}
Eqs.~\eqref{F1}-\eqref{F2} and \eqref{A1}-\eqref{A2} make explicit that, while the perturbative expansion of the influence functional involves full field averages, the perturbative expansion of the influence action exhibits the characteristic \emph{cumulant} structure: $\mathcal{A}_1$ is determined by the mean of the interaction, while $\mathcal{A}_2$ is determined solely by its connected two-point correlator, that is the full two-point average from which the disconnected (factorizable), equal-time contributions have been subtracted.
Since the radiation pressure interaction $S_{\rm int}$ is quadratic in the field variables, we infer from Eqs.~\eqref{A1} and \eqref{A2} that $\mathcal{A}_1$ is composed by equal-time second-order correlators of the field, while $\mathcal{A}_2$ is composed by two-time, fourth-order correlators. For the thermal initial state of the field we consider, the latter decomposes into combinations of quadratic correlators according to Wick's theorem, as outlined in the next section.

\section{Gaussian states and the Wick's decomposition\label{Sec:Wick}}
\subsection{Closed-time-path averages and quantum expectation values\label{Q-CTP-avg}}
The field averages involved in the definition of the influence functional and action can be conveniently computed through functional differentiation of the CTP generating functional for the free field, that we indicate as $Z_{\mathrm{A}}[\{J_{k+}\},\{J_{k-}\}]$. Physically, this represents the influence functional that describes the response of the field linearly coupled to the external sources $J_{k}$ through the field mode amplitudes $Q_k$. According to this interpretation, and similarly to Eq.~\eqref{IF_path}, the free field generating functional is expressed as the CTP average:
\begin{equation}
	Z_{\mathrm{A}}[\{J_{k+}\},\{J_{k-}\}] \equiv \big<e^{i\sum_k(J_{k+}\cdot Q_{k+} - J_{k-}\cdot Q_{k-})}\big>_{\rm A}.
	\label{ZA_3}
\end{equation}
For the sake of readability, we adopt in what follows the compact notation
\begin{equation}
    A\cdot B \equiv \int_{t_i}^{t_f}dt \;A(t)\, B(t),
\end{equation}
to denote the scalar product of the two time-dependent functions $A(t)$ and $B(t)$.

Crucially, a direct correspondence exists between averages evaluated along the CTP contour and quantum expectation values. This correspondence can be established by noting that the free field generating functional admits the following operator representation~\cite{Campos-BR-InIn-1994}:
\begin{align}
    Z_{\mathrm{A}}&[\{J_{k+}\},\{J_{k-}\}] \nonumber\\[4pt]
    &= \text{Tr} \Big\{
\Big[
\widetilde{\mathcal{T}} \exp\Big( -\sum_kiJ_{k-} \cdot \hat Q_k \Big)
\Big]\nonumber\\[4pt]
&\hspace{30pt}\Big[
\mathcal{T} \exp\Big( \sum_k iJ_{k+} \cdot \hat Q_k \Big)
\Big]\,\hat\rho_{\mathrm{A}}(t_i)
\Big\},\label{ZA_2}
\end{align}
where operators are expressed in the Heisenberg picture. Comparing Eqs.~\eqref{ZA_2} and \eqref{ZA_3} we deduce a direct correspondence between time-ordered $\mathcal{T}$ and anti-time-ordered $\widetilde{\mathcal{T}}$ quantum expectation values of field operators and CTP field averages. This relation is expressed explicitly as:
\begin{widetext}
\begin{align}
    &\frac{\partial^{n+m} Z_{\mathrm{A}}}{\partial\big(iJ_{k+}(t_1)\big)\ldots\big(iJ_{j+}(t_n)\big)\;\partial\big(-iJ_{l-}(s_1)\big)\ldots\big(-iJ_{p-}(s_m)\big)}\Bigg|_{\{J_{k\pm}\} =0} = \big\langle {Q}_{j+}(t_n)\ldots{Q}_{k+}(t_1)\;{Q}_{p-}(s_m)\ldots{Q}_{l-}(s_1)\big\rangle_{\mathrm{A}}\nonumber\\[6pt]
    &={\rm Tr}_{\mathrm{A}}\{\widetilde{\mathcal{T}}\big[\hat{Q}_p(s_m)\ldots\hat{Q}_l(s_1)\big]\mathcal{T} \big[\hat{Q}_j(t_n)\ldots\hat{Q}_k(t_1)\big]\hat{\rho}_{\mathrm{A}}(t_i)\}.\label{Q_corr}
\end{align}    
\end{widetext}
Specifically, averages performed along the forward propagating $(+)$ branch of the CTP contour correspond to time-ordered quantum averages, while averages calculated along the backward propagating $(-)$ branch correspond to anti-time-ordered quantum averages. Mixed averages, involving operators on both branches, yield products of anti-time-ordered and time-ordered operators, with the anti-time-ordered block appearing to the left of the time-ordered one, consistent with the operator ordering in Eq.~\eqref{ZA_2}.

\subsection{Free field generating functional and the Wick's decomposition}
For a free field in thermal equilibrium at temperature $T$, the generating functional, as defined in Eq.~\eqref{ZA_3} or \eqref{ZA_2}, can be evaluated exactly, since both the dynamics and the state of the field are quadratic. Employing standard Gaussian path integral techniques, we obtain \cite{QBM1,QBM2,QBM3}:
\begin{equation}
    Z_{\mathrm{A}}[\{J_{k+}\},\{J_{k-}\}] = e^{i\mathcal{W}_{\mathrm{A}}[\{J_{k+}\},\{J_{k-}\}]},\label{Z_Sigma}
\end{equation}
where $\mathcal{W}_{\mathrm{A}}[\{J_{k}^+\},\{J_{k}^-\}]$ is the \emph{connected} generating functional for the free field, which exhibits the characteristic quadratic structure:
\begin{align}
    i\mathcal{W}_{\mathrm{A}}&[\{J_{k}^+\},\{J_{k}^-\}]\nonumber\\[4pt]
    &= -i\sum_k\left(J_{k+} - J_{k-}\right)\cdot (\theta\mu_{k})\cdot\left(J_{k+} + J_{k-}\right) \nonumber\\[0pt]
	&\hspace{12pt}-\sum_k\left(J_{k+} - J_{k-}\right)\cdot (\theta\nu_{k})\cdot \left(J_{k+} - J_{k-}\right). \label{Sigma}
\end{align}
Following a similar notation as the one introduced in the previous section, we used the shorthand:
\begin{equation}
    A\cdot K\cdot B \equiv
    \int_{t_i}^{t_f}dt\int_{t_i}^{t_f}ds \;A(t) \, K(t-s)\, B(s),
\end{equation}
to denote nonlocal time convolutions involving a kernel $K(t)$. 

The connected generating functional for the free field is defined in terms of the modes noise $\nu_k(t)$ and dissipation $\mu_k(t)$ kernels, which take the explicit form:
\begin{subequations}
\begin{align}
	\nu_{k}(t) &= \frac{z_k(T)}{2\omega_{k0}}\cos(\omega_{k0} t),\label{Eq:Nu}\\
	\mu_{k}(t) &= -\frac{1}{2\omega_{k0}}\sin(\omega_{k0} t).\label{Eq:Mu}
\end{align}
\end{subequations}
In the former, both thermal and vacuum fluctuations acting on each mode are encoded within the factor $z_k(T) \equiv 2 n_k(T)+1$, where $n_k(T) = 1/[\exp(\hbar\omega_{k0}/(k_b T))-1]$ is the mode thermal population (with $k_B$ the Boltzmann constant), while the constant $+1$ accounts for the zero-point (vacuum) fluctuations. We highlight that the Heaviside step function $\theta$ has been introduced in Eq.~\eqref{Sigma}, to enforce causal time-ordering in the time integrals. 

\begin{figure}[t]
    \centering
    \includegraphics[width = \linewidth]{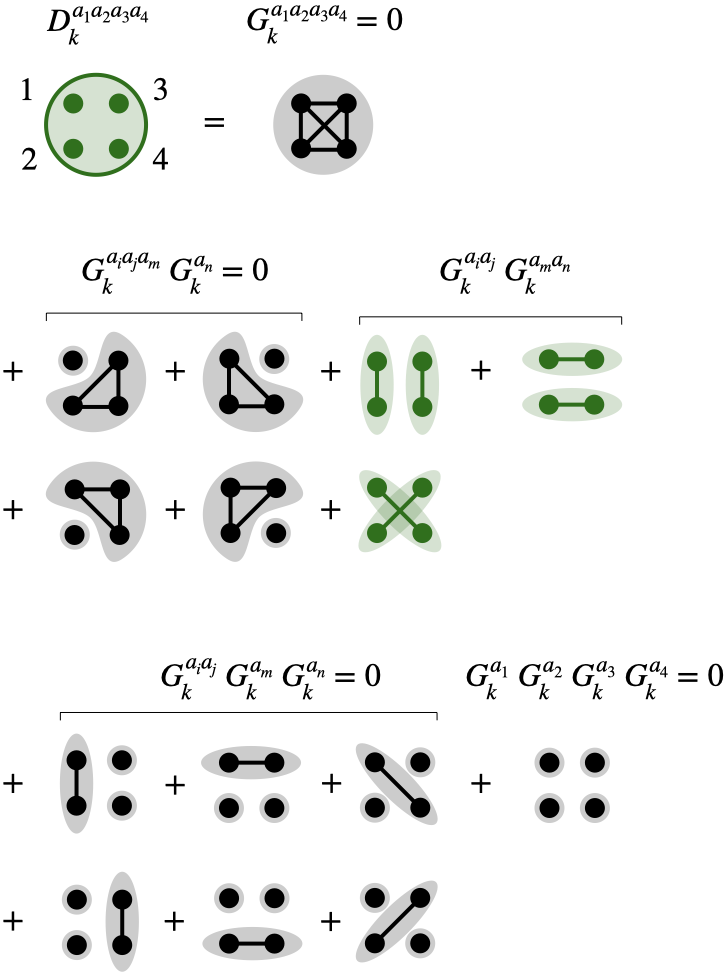}
    \caption{Diagrammatic representation of the decomposition of the fourth-order correlation function $D_k^{a_1 a_2 a_3 a_4}(t_1,t_2,t_3,t_4)$, in terms of connected correlators $G_k^{a_1\ldots a_n}(t_1,\ldots,t_n)$ with $n\leq 4$. The indices $a_i,a_j,a_m,a_n=\pm$ ($i,j,m,n=1,2,3,4$) specify whether the corresponding time arguments lie on the forward $(+)$ or backward $(-)$ branch of the closed-time-path contour. Since the connected generating functional $\mathcal{W}_{\mathrm{A}}$ for the free field in a Gaussian thermal state is quadratic in the sources $J_k$ [see Eq.~\eqref{Sigma}, in the text], all connected correlators with $n\neq 2$ vanish. Therefore, the fourth-order correlation function reduces to a sum of products of two-point closed-time-path propagators, namely the Feynman, Dyson, and Wightman functions.}
\label{Fig:wick_dec}
\end{figure}

Using Eq.~\eqref{Z_Sigma}, any field correlation function:
\begin{multline}
    D_k^{a_1\ldots a_n}(t_1\ldots,t_n)\equiv\langle Q_{ka_1}(t_1)\ldots Q_{ka_n}(t_n)\rangle_{\mathrm{A}} \\
    =\frac{\partial^{n} Z}{\partial\big(ia_1J_{ka_1}(t_1)\big)\ldots\big(ia_nJ_{ka_n}(t_n)\big)}\Bigg|_{J_{k\pm} =0}, \label{Dk}
\end{multline}
with $a_1,\ldots,a_n = \pm$, can be decomposed in terms of functional derivatives of the connected generating functional $\mathcal{W}_{\mathrm{A}}$. These derivatives represent the \emph{connected}, that is not factorizable, correlation functions that we indicated as:
\begin{multline}
    G_k^{a_1\ldots a_n}(t_1\ldots,t_n)\\
    \equiv\frac{\partial^{n} (i\mathcal{W}_{\mathrm{A}})}{\partial\big(ia_1J_{ka_1}(t_1)\big)\ldots\big(ia_nJ_{ka_n}(t_n)\big)}\Bigg|_{J_{k\pm} =0}.\label{G_corr}
\end{multline}
Since $\mathcal{W}_{\mathrm{A}}$ is quadratic in the external sources $J_k$, all such connected correlators with $n\neq 2$ vanish, and the free-field theory is completely characterized by its quadratic correlators. These correspond to the Feynman propagator $G_k^F$, for correlators connecting two points on the forward branch of the CTP contour, the Dyson propagator $G_k^D$, for correlators between points on the backward branch, and the positive- $G_k^{(+)}$ and negative-frequency $G_k^{(-)}$ Wightman functions for correlators mixing the two branches. These fundamental propagators are written in terms of the fluctuation $\nu_k$ and dissipation $\mu_k$ kernels, as:
\begin{align}
    &iG_k^F(t-s) \equiv G_k^{++}(t,s) \nonumber\\[2pt]
    &\qquad=\nu_k(t-s) + i \mu_k(t-s)\;{\rm sign}(t-s),\label{GF}\\[6pt]
    -&iG_k^D(t-s) \equiv G_k^{--}(t,s) \nonumber\\[2pt]
    &\qquad=\nu_k(t-s) - i \mu_k(t-s)\;{\rm sign}(t-s),\label{GD}\\[6pt]
    -&iG_k^{(+)}(t-s) \equiv G_k^{-+}(t,s) \nonumber\\[2pt]
    &\qquad=\nu_k(t-s) + i \mu_k(t-s),\label{G+}\\[6pt]
    -&iG_k^{(-)}(t-s) \equiv G_k^{+-}(t,s) \equiv -iG_k^{(+)}(s-t),\label{G-}
\end{align}
while their operator representation is directly inferred from Eq.~\eqref{Dk}, given the correspondence in Eq.~\eqref{Q_corr}. For a thermal state, any correlation function $D_k^{a_1a_2\ldots a_n}$ of order $n>2$ factorizes into products of these elementary propagators, in accordance with Wick’s theorem. This property is illustrated graphically in Fig.~\ref{Fig:wick_dec}, taking the  fourth-order correlation function $D_k^{a_1a_2a_3a_4}$ as an example. Expanding the fourth-order functional derivative of the generating functional $Z_{\mathrm{A}}$ in terms of functional derivatives of the influence action $\mathcal{W}_{\mathrm{A}}$, the full four-point correlation decomposes as sum of products of connected correlators of order $n\leq 4$. As discussed above, all connected correlators vanish for $n\neq 2$, since the influence functional is quadratic in the sources. Consequently, the four-point correlation function reduces to the following combination of two-point CTP propagators:
\begin{multline}
    D_k^{a_1a_2a_3a_4} = G_k^{a_1a_2}G_k^{a_3a_4} + G_k^{a_1a_3}G_k^{a_2a_4} \\+ G_k^{a_1a_4}G_k^{a_2a_3}.
\end{multline}
The diagrammatic representation of the Wick's decomposition that leads to this result is given in Fig.~\eqref{Fig:wick_dec}.

\section{Influence action of the radiation pressure interaction\label{sec:rad_pres_infl_func}}

After the preparatory work carried out in the previous sections, we are now equipped with the theoretical tools that we need to derive the effective action for the moving mirror including full back-reaction of the cavity field. Through the calculation of the influence action for the mirror, we show that back-reaction effects appear in the form of colored quantum noise, and corresponding nonlocal dissipation, generated both by the stimulated and spontaneous photons creation in the cavity induced by the mechanical motion. The stimulated emission results from the parametric excitation of particles already present in the initial thermal state of the field, while the spontaneous emission is seeded by field vacuum fluctuations. Noise and dissipation are captured by corresponding fluctuation and dissipation kernels, whose derivation is the main objective of this section.

Due to the nonlinear structure of the radiation pressure interaction, we pursue the perturbative approach outlined in Sec.~\ref{sec:pert_IA}. Such a perturbative treatment is justified provided the radiation pressure interaction constitutes a small correction to the free dynamics of the combined mirror-field system. This condition is verified within standard optomechanical regimes, as we discuss in Sec.~\ref{sec:pert}.

In the following, our goal is to determine both the first-order $\mathcal{A}_1$ and the second-order $\mathcal{A}_2$ components of the mirror effective action, as defined in Eqs.~\eqref{A1} and~\eqref{A2}, respectively. In Sec.~\ref{Sec:A1}, we show that the first-order term gives rise to static effects that arise due to the cavity-length dependence of the field modes frequencies, and yields the static Casimir potential. The second-order term, which encodes the nonlinear fluctuation and dissipation kernels governing the dynamical back-reaction of the cavity field on the mirror, is derived and discussed in details in Sec.~\ref{Sec:A2}.

\subsection{Small displacement limit and the perturbative parameter\label{sec:pert}}

We work within the usual assumption of small mechanical displacements $x$ of the mirror relative to the equilibrium length $d$ of the cavity. This assumption permits us to use the relative displacement $y \equiv x/d \ll 1$ as the small perturbative parameter to expand the interaction. Writing the mirror coordinate as $q(t) = d + x(t) = d[1+y(t)]$, and expanding up to second order in $y$, each of the interaction terms defined in Eqs.~\eqref{Sint_0}-\eqref{Sint_2} read as:
\begin{subequations}
\begin{align}
    S_{\rm int}^{(0)}&[y,\{Q_k\}] = \int_{t_i}^{t_f} dt \,y(t)
 \left(\sum_{k=1}^{+\infty} \omega_{k0}^2 \, Q_k^2(t)\right) \nonumber\\[2pt]
&- \frac{3}{2} \int_{t_i}^{t_f} dt \, y^2(t)
\left( \sum_{k=1}^{+\infty} \omega_{k0}^2 \, Q_k^2(t) \right)
 \label{Sint_x_0},\\[6pt]
    S_{\rm int}^{(1)}&[y,\{Q_k\}] = \int_{t_i}^{t_f} dt \, \dot{y}(t)
\left( \sum_{k=1}^{+\infty} \dot{Q}_k(t) \, Q_k^{(1)} (t)\right)\nonumber\\[2pt]
&- \int_{t_i}^{t_f} dt \, \dot{y}(t)\, y(t)
\left( \sum_{k=1}^{+\infty} \dot{Q}_k(t) \, Q_k^{(1)}(t) \right)
,\label{Sint_x_1}\\[6pt]
    S_{\rm int}^{(2)}&[y,\{Q_k\}] \nonumber\\ &= - \frac{1}{2} \int_{t_i}^{t_f} dt \, \dot{y}^2(t) \left(\sum_{k=1}^{+\infty} Q_k(t) \, Q_k^{(2)}(t)\right).\label{Sint_x_2}
\end{align}
\end{subequations}
These expansions reveal that both $S_{\mathrm{int}}^{(0)}$ and $S_{\mathrm{int}}^{(1)}$ contain contributions linear in $y$ (or the velocity $\dot{y}$), while $S_{\mathrm{int}}^{(2)}$ is already quadratic in these quantities. Consequently, when expanding the influence action to second order in the relative displacement, all three interaction terms contribute, at least formally, to the first-order component $\mathcal{A}_1$, whereas only $S_{\mathrm{int}}^{(0)}$ and $S_{\mathrm{int}}^{(1)}$ contribute to the second-order component $\mathcal{A}_2$, the latter being by definition quadratic in the interaction.

Before proceeding, let us spend a few more words on the perturbative parameter underlying the expansion of the radiation pressure interaction. Strictly speaking, the perturbative parameter of the theory is not simply the relative displacement $y$, but rather the ratio between the strength of the interaction and the characteristic energy scale of the unperturbed system. We take the quantum of mechanical energy $\hbar\Omega$, with $\Omega$ the characteristic oscillation frequency of the mirror, as a reference for the latter. The strength of the interaction can be estimated by studying the Lagrangian corresponding to the lowest order term of the interaction, namely the first term of Eq.~\eqref{Sint_x_0}: $L_{\mathrm{int}}^{(0)}[y,\{Q_k\}]\equiv \sum_k\omega_{k0}^2\,y(t)Q_k^2(t)$ . As with any other interaction term, this is formally ultraviolet divergent due to the summations over the infinitely many field modes accounted for by the radiation pressure model. This pathological behaviour is regularized on physical grounds, since field modes with frequencies higher than the plasma frequency $\omega_{\mathrm{pl}}$ of the mirror material cannot effectively couple to the mirror itself. Performing an order-of-magnitude analysis, we take the plasma frequency as a physically justified cut-off for sums and consider the zero-temperature limit of the theory for ease. Under these assumptions, the interaction strength is, to lowest order~\cite{Butera-2025-corrections}:
\begin{equation}
    \langle  L_{\rm int}^{(0)}[y,\{Q_k\}] \rangle_{\mathrm{A}} \sim ({x_{\mathrm{zpf}}}/{d})\,\hbar\omega_{\mathrm{pl}}. \label{int_strength}
\end{equation}
Here, $x_{\rm zpf} = [{\hbar/(2m\Omega)}]^{1/2}$ is the amplitude of the vacuum fluctuations of the mirror position within the confining harmonic potential. The perturbative parameter of the theory is then obtained as the ratio between Eq.~\eqref{int_strength} and $\hbar\Omega$:
\begin{equation}
  \lambda = \frac{x_{\mathrm{zpf}}}{d} \frac{\omega_{\rm pl}}{\Omega}.
\end{equation}
This parameter measures for the relative strength of the single-photon radiation pressure interaction compared to the unperturbed mechanical energy. Except in the so-called \emph{ultra-strong coupling} regime~\cite{UltraStrong_Review}, where $\lambda \approx 1$, in both the \emph{weak coupling} and \emph{strong coupling} limits it is verified that $\lambda \ll 1$, and the perturbative treatment of the radiation pressure interaction is therefore permitted. Having clarified this point, we shall continue to refer to $y$ as the perturbative parameter in what follows, for simplicity.

\subsection{Influence action: first-order term\label{Sec:A1}}

We begin with the evaluation of the first-order term $\mathcal{A}_1$ of the influence action. This is composed by three different contributions, one for each of the interaction components $S_{\mathrm{int}}^{(n)}$, with $n=0,\, 1,\, 2$. We pursue the following calculations by using the small displacement expansions in Eqs.~\eqref{Sint_x_0}-\eqref{Sint_x_2}, retaining terms up to second order in the relative displacement $y$. Each interaction component formally contribute with a term of the form:
\begin{equation}
    \mathcal{A}_1^{(n)}[y_+,y_-] = \langle S_{\rm int}^{(n)}[+] \rangle_{\mathrm{A}} - \langle S_{\rm int}^{(n)}[-]\rangle_{\mathrm{A}}. \label{A_1_n}
\end{equation}
Among these components, it turns out that only $\mathcal{A}_1^{(0)}$ is relevant and gives rise to the static Casimir potential. The term $\mathcal{A}_1^{(1)}$, which is linear in the mirror velocity, vanishes identically, while $\mathcal{A}_1^{(2)}$, which is quadratic in the mirror velocity, is exactly cancelled by an equal and opposite component arising from the second-order term $\mathcal{A}_2$ of the influence action, as discussed in Sec.~\ref{Sec:A2}.

\subsubsection{Position-dependent component: the Casimir potential}

Let us start with the calculation of the first-order component purely depending on the position of the mirror. This is readily obtained by substituting Eq.~\eqref{Sint_x_0} in Eq.~\eqref{A_1_n}, with $n = 0$. Using the equal-time field averages
\begin{equation}
     \langle Q_{ka}^2(t)\rangle_{\mathrm{A}} = G_{k}^{aa}(t,t) = \nu_k(0),
\end{equation}
with $a =\pm$, and given $\nu_k(0) = z_k(T)/(2\omega_{k0})$ [see Eq.~\eqref{Eq:Nu}], we obtain:
\begin{multline}
    \mathcal{A}_1^{(0)}[y_+,y_-] \\
    = \frac{E_{\mathrm{A}}(T)}{2}\int_{t_i}^{t_f}\!dt
    \left[(y_+ - y_-) - \frac{3}{2}(y_+^2 - y_-^2)\right],
    \label{A1_0}
\end{multline}
where
\begin{equation}
    E_{\mathrm{A}}(T) = \sum_{k=1}^{+\infty} z_k(T)\,\frac{\omega_{k0}}{2}
    \label{EA}
\end{equation}
is the total energy stored in the cavity field at temperature $T$. This static quantity is ultraviolet divergence but can be renormalized by subtracting the corresponding free-space value obtained in the limit of infinite cavity length $d \to +\infty$. More precisely, the renormalization procedure should be performed on the field energy density $\varepsilon_{\mathrm{A}} \equiv E_{\mathrm{A}}/d$, which is an intensive quantity, as opposed to the energy, that is an extensive quantity and therefore carries an implicit dependence on the geometry of the system.

The renormalization procedure starts by regularizing the divergent energy density through the introduction of a frequency cut-off $\sigma^{-1}$ in the sum over the field modes:
\begin{equation}
	\varepsilon_{\mathrm{A}}^{\rm reg} \equiv \sum_{k=1}^{+\infty} \;z_k\frac{\omega_{k0}}{2d} \exp{(-\sigma\omega_{k0})}. \label{reg_en_dens}
\end{equation}
The renormalized energy density is then obtained by subtracting the divergent free-space value, i.e., the $d\to+\infty$ limit, and subsequently removing the regulator:
\begin{equation}
	\varepsilon_{\mathrm{A}}^{\rm ren} \equiv \lim_{\sigma\to 0}\left(\varepsilon_{\mathrm{A}}^{\rm reg} - \lim_{d\to +\infty}\varepsilon_{\mathrm{A}}^{\rm reg}\right).
\end{equation}
This renormalized quantity determines the physical energy density entering the influence action:
\begin{multline}
    \mathcal{A}_{1,\rm ren}^{(0)} [x_+,x_-]  \\
     = \varepsilon_{\mathrm{A}}^{\rm ren}(T)\int_{t_i}^{t_f} \!dt  \left[ (x_+-x_-) - \frac{3}{2d}(x_+^2-x_-^2)\right].\label{A01_ren}
\end{multline}
This action provides the expansion of the Casimir energy to quadratic order in the mirror displacement $x$, which gives rise to the Casimir force between two parallel mirrors in the one-dimensional configuration considered here. Below, we verify this by examining the zero-temperature and high-temperature limits.

At zero-temperature, the field modes are populated solely by vacuum fluctuations, so that $z_k\to 1$. The presence of the high-frequency cutoff allows the mode sum in Eq.~\eqref{reg_en_dens} to be
evaluated explicitly, yielding the regularized expression
\begin{equation}
	\varepsilon_{\mathrm{A}}^{\rm reg} (T=0)  = \frac{\hbar}{2\pi\sigma^2} - \frac{\hbar\pi}{24d^2} + \mathcal{O}(\sigma^2).
\label{reg_en_dens_T0}
\end{equation}
The first term is the universal, geometry-independent vacuum divergence, which persists in the free-space limit $d\to+\infty$. Subtracting it from Eq.~\eqref{reg_en_dens_T0} and removing the regulator yields the renormalized zero-temperature energy density
\begin{equation}
    \varepsilon_{\mathrm{A}}^{\rm ren} (T=0)  = - \frac{\pi}{24 d^2}.\label{ren_en_dens_T0}
\end{equation}
This is the characteristic Casimir energy density for a one-dimensional system of length $d$~\cite{Law-MirFieldInt-1995}.

The same procedure applies in the opposite, high-temperature limit. In this case, we first regularize the energy density with the cut-off frequency $\sigma^{-1}$, and then assume ${k_B T}/{\hbar\sigma^{-1}}\gg 1$. In such a limit, $z_k\to {2k_BT}/{\hbar\omega_{k0}}$, and the regularized energy density writes:
\begin{equation}
	\varepsilon_{\mathrm{A}}^{\rm reg} \left(\frac{k_BT}{\hbar\omega_{k0}}\gg 1 \right)  = k_BT \left(\frac{1}{\pi\sigma} + \frac{1}{2d}+ \mathcal{O}(\sigma)\right).
\label{reg_en_dens_highT}
\end{equation}
Identifying the first term as the universal vacuum divergence and subtracting it, the renormalized energy density in the high-temperature limit take the expected form
\begin{equation}
    \varepsilon_{\mathrm{A}}^{\rm ren} \left(\frac{k_BT}{\hbar\omega_{k0}}\gg 1 \right)  = \frac{k_B T}{2 d},
\end{equation}
which is linear in the thermal energy.

\subsubsection{Component linear in the mirror velocity}
The component of the first-order influence action that is linear in the mirror velocity vanishes identically. In fact, specializing Eq.~\eqref{A_1_n} to $n = 1$, we
finds that this component is proportional to the field correlators
\begin{equation}
    \langle \dot{Q}_{ka}\, Q_{ka}^{(1)}\rangle_{\mathrm{A}} =
    \sum_j g_{jk}\langle \dot{Q}_{ka}\, Q_{ja}\rangle_{\mathrm{A}} = 0,
\end{equation}
with $a = \pm$. This sum vanishes identically because the different field modes are assumed to be initially uncorrelated and the coupling coefficients satisfy $g_{kk} = 0$ by construction [see Eq.~\eqref{g_kj}].

\subsubsection{Component quadratic in the mirror velocity\label{Sec:A1_2}}

For completeness, we give the expression for the component of the first-order influence action that is quadratic in the mirror velocity. As noted above, this term does not contribute to the final expression of the influence action, since an equal and opposite term arises from the second-order term $\mathcal{A}_2$. Specializing Eq.~\eqref{A_1_n} to $n = 2$ and using Eqs.~\eqref{Qn}, \eqref{GF}, and~\eqref{GD}, this term reads
\begin{equation}
    \mathcal{A}_1^{(2)}[y_+,y_-] = \frac{\Delta m}{2}\int_{t_i}^{t_f}dt
    \left(\dot{y}_+^2 - \dot{y}_-^2\right),
\end{equation}
with the mass shift:
\begin{equation}
    \Delta m = \sum_{kj=1}^{+\infty} g_{kj}^2 \,\nu_k(0).
\end{equation}

\subsection{Influence action: second-order term\label{Sec:A2}}

The calculation of the second-order term of the influence action is more involved, as it requires the evaluation of higher-order two-point correlation functions of the field. Using the definition in Eq.~\eqref{A2} together with the expansions in Eqs.~\eqref{Sint_x_0}-\eqref{Sint_x_2}, $\mathcal{A}_2$ can be written as a sum of the following different contributions:
\begin{equation}
    \mathcal{A}_{2}[y_+,y_-] = \frac{i}{2}\sum_{i,j=0}^2\mathcal{A}_{2}^{ij}[y_+,y_-],\label{A2sum}
\end{equation}
with each term defined as:
\begin{equation}
    \mathcal{A}_{2}^{ij}[y_+,y_-] =\sum_{a,b=\pm} a b\,\big\langle S_{\rm int}^{(i)}[a] S_{\rm int}^{(j)}[b] \big\rangle_{\bar{A}}.\label{A2_ij}
\end{equation}
Here $\langle...\rangle_{\bar{A}}$ denotes the nonlocal component of the field averages, from which the same-time contribution has been subtracted:
\begin{multline}
        \big\langle S_{\rm int}^{(i)}[a] {S}_{\rm int}^{(j)} [b]\big\rangle_{\bar{A}}
        \equiv \big\langle S_{\rm int}^{(i)}[a] {S}_{\rm int}^{(j)} [b]\big\rangle_{\mathrm{A}} \\
        - \big\langle S_{\rm int}^{(i)}[a]\big\rangle_{\mathrm{A}} \big\langle{S}_{\rm int}^{(j)} [b]\big\rangle_{\mathrm{A}}. \label{connected}
\end{multline}
Since the radiation pressure interaction is quadratic in the field variables, these terms involve fourth-order correlators and their derivatives, that can be decomposed into products of quadratic correlators via Wick's theorem, as presented in Sec.~\ref{Sec:Wick}.

Since the theory is developed to quadratic order in the mirror relative displacement $y$ and velocity $\dot{y}$, not all terms $\mathcal{A}_2^{ij}$ contribute. Specifically, since $S_{\mathrm{int}}^{(0)}$ and $S_{\mathrm{int}}^{(1)}$ are linear in these quantities while $S_{\mathrm{int}}^{(2)}$ is quadratic, the
terms $\mathcal{A}_2^{i2}$ with $i = 0,1,2$ are of third order and can be discarded. Quadratic contributions arise only from $\mathcal{A}_2^{00}$, $\mathcal{A}_2^{01}+\mathcal{A}_2^{10}$, and $\mathcal{A}_2^{11}$, whose Wick decompositions are discussed in the next section.

\subsubsection{Wick's decomposition of quartic correlators}

The correlation functions that contribute to the influence action components $\mathcal{A}_2^{00}$, $\mathcal{A}_2^{01}$ $(\mathcal{A}_2^{10})$ and $\mathcal{A}_2^{11}$ are defined in terms of fourth-order field correlators. They take the explicit form:
\begin{multline}
    \big\langle S_{\mathrm{int}}^{(0)}[a]\,S_{\mathrm{int}}^{(0)}[b]
    \big\rangle_{\bar{A}} = \sum_{k,j=1}^{+\infty}\omega_{k0}^2\,\omega_{j0}^2
    \int_{t_i}^{t_f}\!dt \int_{t_i}^{t_f}\!ds \\[4pt]
    \times y_a(t)\,\big\langle Q_{ka}(t)Q_{ka}(t)\,
    Q_{jb}(s)Q_{jb}(s)\big\rangle_{\bar{A}}\,y_b(s), \label{S0S0_ab}
\end{multline}

\begin{multline}
    \big\langle S_{\mathrm{int}}^{(1)}[a]\,S_{\mathrm{int}}^{(1)}[b]
    \big\rangle_{\bar{A}} = \sum_{k,j=1}^{+\infty}
    \int_{t_i}^{t_f}\!dt \int_{t_i}^{t_f}\!ds\\[4pt]
    \times\dot{y}_a(t)\,\big\langle \dot{Q}_{ka}(t)\,Q_{ka}^{(1)}(t)\,
    \dot{Q}_{jb}(s)\,Q_{jb}^{(1)}(s)\big\rangle_{\bar{A}}\,
    \dot{y}_b(s), \label{S1S1_ab}
\end{multline}

\begin{multline}
    \big\langle S_{\mathrm{int}}^{(0)}[a]\,S_{\mathrm{int}}^{(1)}[b]
    \big\rangle_{\bar{A}} = \sum_{k,j=1}^{+\infty}\omega_{k0}^2
    \int_{t_i}^{t_f}\!dt \int_{t_i}^{t_f}\!ds\\[4pt]
    \times y_a(t)\,\big\langle Q_{ka}(t)Q_{ka}(t)\,
    \dot{Q}_{jb}(s)\,Q_{jb}^{(1)}(s)\big\rangle_{\bar{A}}\,
    \dot{y}_b(s), \label{S0S1_ab}
\end{multline}
where as before $a,b = \pm$ specify whether the time arguments lie on the forward $(+)$ or backward $(-)$ branch of the CTP contour. Since the field state is Gaussian, these fourth-order correlators can be decomposed into products of quadratic ones via Wick's theorem:
\begin{align}
\langle Q_{ka}&(t)Q_{ka}(t)Q_{jb}(s)Q_{jb}(s)\,\rangle_{\bar{A}} \nonumber\\[2pt]
&\qquad =2\langle Q_{ka}(t)\,Q_{jb}(s)\rangle_{\mathrm{A}}^2,\label{wick_00}\\[6pt]
\langle\dot{Q}_{ka}&(t) Q_{ka}^{(1)}(t) \dot{Q}_{jb}(s) Q_{jb}^{(1)}(s)
\rangle_{\bar{A}}\nonumber\\[2pt]
&\qquad = \big\langle \dot{Q}_{ka}(t) \dot{Q}_{jb}(s) \big\rangle_{\mathrm{A}} 
\big\langle Q_{ka}^{(1)}(t) Q_{jb}^{(1)}(s) \big\rangle_{\mathrm{A}}\nonumber\\[2pt]
&\qquad+\big\langle \dot{Q}_{ka}(t) Q_{jb}^{(1)}(s) \big\rangle_{\mathrm{A}}
\big\langle  Q_{ka}^{(1)}(t) \dot{Q}_{jb}(s)  \big\rangle_{\mathrm{A}},\label{wick_11}\\[6pt]
\langle Q_{ka}&(t) Q_{ka}(t) \dot{Q}_{jb}(s) Q_{jb}^{(1)}(s)\rangle_{\hat{A}}\nonumber\\[2pt]
 & \qquad= 2\big\langle {Q}_{ka}(t) \dot Q_{jb}(s) \big\rangle_{\mathrm{A}}
\big\langle  Q_{ka}(t) {Q}_{jb}^{(1)}(s)  \big\rangle_{\mathrm{A}}\label{wick_01}.
\end{align}
Substituting Eqs.~\eqref{wick_00}-\eqref{wick_11} into
Eqs.~\eqref{S0S0_ab}-\eqref{S1S1_ab} yields
\begin{multline}
    \big\langle S_{\rm int}^{(0)}[a] S_{\rm int}^{(0)}[b] \big\rangle_{\hat{A}} = 2\sum_{k=1}^{+\infty}\omega_{k0}^4\int_{t_i}^{t_f}\!dt\int_{t_i}^{t_f}\!ds\\
     \times y_a(t)\,\big[G_k^{ab}(t-s)\big]^2 \;y_b(s) \label{S0S0}
\end{multline}
and
\begin{align}
    \big\langle& S_{\mathrm{int}}^{(1)}[a]\,S_{\mathrm{int}}^{(1)}[b]
    \big\rangle_{\bar{A}} = \sum_{k,j=1}^{+\infty}
    \int_{t_i}^{t_f}\!dt\int_{t_i}^{t_f}\!ds\;
    \nonumber\\[4pt]
    &\times\dot{y}_a(t)\bigg\{
    \big[\partial_t\partial_s G_k^{ab}(t-s)\big]
    \big[g_{jk}^2\,G_j^{ab}(t-s)\big] \nonumber\\[4pt]
    &\phantom{{}\times{}}+
    \big[g_{kj}\,\partial_t G_k^{ab}(t-s)\big]
    \big[g_{jk}\,\partial_s G_j^{ab}(t-s)\big]
    \bigg\}\dot{y}_b(s). \label{S1S1}
\end{align}
The term mixing $S_{\rm int}^{(0)}$ and $S_{\rm int}^{(1)}$ vanishes identically:
\begin{align}
    \big\langle& S_{\rm int}^{(0)}[a] S_{\rm int}^{(1)}[b] \big\rangle_{\hat{A}} \nonumber\\[4pt]
    &=2\sum_{k=1}^{+\infty} \,\omega_{k0}^2\, \, y_a(t) \,\big[\partial_s G_k^{ab}(t-s)\big]\big[g_{kk}G_k^{ab}(t-s)\big] \,\dot{y}_b(s) \nonumber\\[4pt] & =\; 0,
\end{align}
since $g_{kk}=0$ by construction [see Eq.~\eqref{g_kj}]. Consequently, both $\mathcal{A}_{2}^{01}[y_+,y_-]$ and $\mathcal{A}_{2}^{10}[y_+,y_-]$ vanish, and the second-order term of the influence action receives contributions
only from:
\begin{align}
    \mathcal{A}_{2}[y_+,y_-] = \mathcal{A}_{2}^{00}[y_+,y_-] + \mathcal{A}_{2}^{11}[y_+,y_-].
\end{align}
In the following section, we discuss the structure of these two terms and show that they introduce the fluctuation and dissipation kernels encoding the full dynamical back-reaction from the field in the form of  noise and dissipation acting on the mirror. We study the physical origin of these back-reaction effects, showing that they relate to nonequilibrium particle creation processes in the field by DCE.

\subsubsection{Effective fluctuation and dissipation kernels}
The explicit expressions of $\mathcal{A}_{2}^{00}$ and $\mathcal{A}_{2}^{11}$ are obtained by specializing Eq.~\eqref{A2_ij} using Eqs.~\eqref{S0S0} and~\eqref{S1S1}, together with the expressions for the field propagators in terms of the single-mode fluctuation and dissipation kernels, Eqs.~\eqref{GF}-\eqref{G-}. After lengthy but standard closed-time-path integral calculations (see, e.g., \cite{QBM1,QBM2,QBM3}), we obtain:
\begin{widetext}
\begin{multline}
    \mathcal{A}_{2}^{00}[y_+,y_-] = 2\, i 
\int_{t_{i}}^{t_{f}} d t
\int_{t_{i}}^{t} d s
\Big\{
\big[y_+(t) - y_-(t)\big]\,
\big[N_+^{00}(t-s) + N_-^{00}(t-s)\big]\,
\big[y_+(s) - y_-(s)\big] \\
+i\big[y_+(t) - y_-(t)\big]M_+^{00}(t - s)\,
\big[y_+(s) + y_-(s)\big]\Big\},\label{A_2_00}
\end{multline}
and
\begin{multline}
    \mathcal{A}_{2}^{11}[y_+,y_-] =  \frac{i}{2} 
\int_{t_{i}}^{t_{f}} d t
\int_{t_{i}}^{t} d s
\Big\{
\big[\dot y_+(t) - \dot y_-(t)\big]\,
\,\big[N_{+}^{11}(t - s)+N_{-}^{11}(t - s)\big]
\big[\dot y_+(s) - \dot y_-(s)\big] \\
+i\big[\dot y_+(t) - \dot y_-(t)\big]\,
 \big[M_{+}^{11}(t - s) + M_{-}^{11}(t - s)\big]\,
\big[\dot y_+(s) + \dot y_-(s)\big]\Big\}  -  \frac{\Delta m}{2} \int_{t_i}^{t_f} dt   \left( \dot{y}_+^2 
- \dot{y}_-^2 \right).\label{A_2_11}
\end{multline}
\end{widetext}
Colored noise and nonlocal dissipation acting on the mirror are encoded in the fluctuation kernels $N_\pm^{00}(t)$, $N_\pm^{11}(t)$ and the dissipation kernels $M_+^{00}(t)$, $M_\pm^{11}(t)$, defined as
\begin{subequations}
\begin{align}
    N_\pm^{00}(t) &= \sum_{k=1}^{+\infty} \;\omega_{k0}^{2} \;\nu_{\pm}^{(2)}[t;k,k], \label{N00}\\
    M_+^{00}(t) &= \sum_{k=1}^{+\infty} \;\omega_{k0}^{2}\;\mu_{+}^{(2)}[t;k,k],\label{M00}
\end{align}
\end{subequations}
and
\begin{subequations}
\begin{align}
    N_\pm^{11}(t) &= \sum_{k,j=1}^{+\infty}  g_{kj}^{2} \;
 \frac{(\omega_{k0} \mp \omega_{j0})^{2}}{\omega_{k0}\,\omega_{j0}} 
\, \nu_{\pm}^{(2)}[t ;k , j],\label{N11}\\
    M_\pm^{11}(t) &= \sum_{k,j=1}^{+\infty}  g_{kj}^{2} \;
 \frac{(\omega_{k0} \mp \omega_{j0})^{2}}{\omega_{k0}\,\omega_{j0}} 
\, \mu_{\pm}^{(2)}[t ; k , j].\label{M11}
\end{align}
\end{subequations}
The structures of both sets of kernels displays dependence on the microscopic fluctuation kernel $\nu_\pm(t)$ and dissipation kernel $\mu_\pm(t)$, defined as:
\begin{subequations}
\begin{align}
    \nu_{\pm}^{(2)}(t \,;\, k , j)
&= \tfrac{1}{8}\, (z_{k}z_{j} \pm 1)\,
\cos\!\big[(\omega_{k} \pm \omega_{j})\, t\big],\label{nu_2}\\
\mu_{\pm}^{(2)}(t \,;\, k , j)
&= \mp \tfrac{1}{8}\, (z_{k} \pm z_{j})\,
\sin\!\big[(\omega_{k} \pm \omega_{j})\, t\big],\label{mu_2}
\end{align}
\end{subequations}
which arise from the coupling between the mirror and pairs of field modes. This coupling is conveyed through two different channels, labeled $(+)$ and $(-)$, depending on whether the mode frequencies add or subtract in the interaction. Note that the contribution $N_-^{00}$ appearing in Eq.~\eqref{A_2_00} is static (zero-frequency) and therefore carries no dissipative counterpart, consistent with the fluctuation-dissipation theorem. Such a term vanishes in the zero-temperature limit. As anticipated, the last term in Eq.~\eqref{A_2_11} is a mass shift that exactly cancels the equal and opposite contribution arising from the first-order component $\mathcal{A}_1$ of the effective action.

Beyond their common dependence on the same microscopic kernels, the relationship between $N_\pm^{00}$, $M_+^{00}$ and $N_\pm^{11}$, $M_\pm^{11}$ is not immediately transparent in the time domain. This connection becomes explicit in the frequency domain, as discussed in the following section, where we demonstrate that the $00$ and $11$ kernels are intimately related, as both account for the same underlying physical process, namely the emission of photon pairs in the cavity. Specifically, the $00$ and $11$ kernels describe intra-mode and inter-mode photon pair creation, respectively, that is the emission of couples of photons either within the same or different field modes. This picture connects directly to the nonequilibrium quantum field mechanism responsible for the back-reaction, namely the DCE.

\subsection{Spectral content of the influence action\label{A_spectr}}

By opportunely rearranging and combining Eqs.~\eqref{N00}-\eqref{M00} and Eqs.~\eqref{N11}-\eqref{M11}, using the definitions given above for the microscopic kernels $\nu_{\pm}^{(2)}$ and $\mu_{\pm}^{(2)}$, we find that the mirror fluctuation and dissipation kernels admit the following discrete spectral decompositions:
\begin{subequations}
\begin{align}
    N_+^{00}(t) &= \sum_{k=-\infty}^{+\infty} \;\omega_{k}^2\,\nu_{kk}^{(2)} \;e^{2i\omega_k t},\label{N00_spectr}\\
    M_+^{00}(t) &=  \sum_{k=-\infty}^{+\infty} \;\omega_{k}^2\,\mu_{kk}^{(2)} \;e^{2i\omega_k t},\label{M00_spectr}
\end{align}
\end{subequations}
and
\begin{subequations}
\begin{align}
    N_{+}^{11}(t)+N_{-}^{11}(t) &=\sum_{\substack{k,j=-\infty \\ k\neq j}}^{+\infty} \nu_{kj}^{(2)} e^{i(\omega_k+\omega_j) t},\label{N11_spectr}\\[2pt]
     M_{+}^{11}(t)+M_{-}^{11}(t) &= \sum_{\substack{k,j=-\infty \\ k\neq j}}^{+\infty} \mu_{kj}^{(2)} e^{i(\omega_k+\omega_j) t},\label{M11_spectr}
\end{align}
\end{subequations}
where 
\begin{subequations}
\begin{align}
    \nu_{kj}^{(2)} &= \frac{\omega_k \omega_j}{4(\omega_k+\omega_j)^2}(z_k z_j +1),\\
    \mu_{kj}^{(2)} &= i\frac{\omega_k \omega_j}{4(\omega_k+\omega_j)^2}(z_k+z_j),
\end{align}
\end{subequations}
are the corresponding discrete spectral components. As expected, these satisfy a fluctuation-dissipation relation of the form
\begin{equation}
    \nu_{kj}^{(2)} = - i \frac{z_k z_j + 1}{z_k + z_j} \,\mu_{kj}^{(2)} = - i z_{k+j} \,\mu_{kj}^{(2)},
\end{equation}
where the factor $z_{k+j} \equiv \coth[\hbar(\omega_k+\omega_j)/2k_BT]$ encodes both thermal and vacuum fluctuations evaluated at the sum frequency $\omega_k+\omega_j$. These results show that the $00$ and $11$ kernels share the same spectral structure, both depending on $\nu_{kj}^{(2)}$ and $\mu_{kj}^{(2)}$ evaluated at the sum frequency $\omega_k + \omega_j$. This common structure reflects the fact that fluctuation and dissipation experienced by the mirror originate from its interaction with photon pairs, with the $00$ kernels accounting for the coupling with photon pairs within the same field mode ($k = j$) and the $11$ kernels accounting for the coupling with pairs across distinct modes ($k \neq j$).

Defining the Fourier transform of the generic time-dependent function $f(t)$ as:
\begin{equation}
    \tilde{f}_\omega = \frac{1}{2\pi}\int_{-\infty}^{+\infty} dt\,f(t)\,e^{-i\omega t},\label{FT}
\end{equation}
with inverse
\begin{equation}
    f(t) = \int_{-\infty}^{+\infty} d\omega\,\tilde{f}_\omega\,e^{i\omega t},\label{Inv_FT}
\end{equation}
and using the spectral representations of the kernels given above, the
second-order influence action admits the following discrete frequency-domain
representation:
\begin{align}
    \mathcal{A}_{2}&[\tilde x_+, \tilde x_-] = \mathcal{A}_{2}^{00}[\tilde x_+,\tilde x_-] + \mathcal{A}_{2}^{11}[\tilde x_+,\tilde x_-]\nonumber\\
    &=\frac{i\pi^2}{d^2}\sum_{k,j=-\infty}^{+\infty}(\omega_k+\omega_j)^2\,\Big(\nu_{kj}^{(2)}+ i \mu_{kj}^{(2)}\Big) \,|\tilde x_+-\tilde x_-|^2_{\omega_k+\omega_j}\nonumber\\
    &+\frac{i\pi^2}{d^2}\sum_{k=-\infty}^{+\infty}(2\omega_k)^2\,\nu_{kk}^{(2)}\, |\tilde x_+-\tilde x_-|_{0}^2.\label{A_FT_discr}
\end{align}
This result can be conveniently recast by taking the continuum limit of infinite cavity length, $d\to+\infty$. In this limit, discrete sums over field modes are replaced by continuous integrals over frequency, according to the prescription
\begin{equation}
    \sum_{k=-\infty}^{+\infty} \to\frac{d}{\pi}\int_{-\infty}^{+\infty}d\omega,
\end{equation}
where the prefactor $d/\pi$ is the one-dimensional density of states. The last term in Eq.~\eqref{A_FT_discr} vanishes in this limit (once the diverging sum is regularized through the plasma frequency cut-off), and the Fourier representation of the influence action reduces to
\begin{multline}
    \mathcal{A}_{2}[\tilde x_+, \tilde x_-] =i \int_{-\infty}^{+\infty}d\omega \int_{-\infty}^{+\infty}d\omega'\\
    \times(\omega+\omega')^2\,\Big(\nu_{\omega\omega'}^{(2)}+ i \mu_{\omega\omega'}^{(2)}\Big) \,|\tilde x_+-\tilde x_-|^2_{\omega+\omega'}.\label{A_FT}
\end{multline}
This result suggests the following physical interpretation: The radiation pressure interaction couples the moving mirror to pairs of field modes with frequencies $\omega$ and $\omega'$. Due to this interaction the mirror experiences fluctuation and corresponding dissipative forces at the sum frequency $\omega + \omega'$ resonant with these mode pairs. As mentioned in the previous section, this property hints that the back-reaction of the field on the mirror dynamics originates from the emission of photon pairs into these modes, that is a direct signature of DCE emission. This interpretation will be further substantiated in Secs.~\ref{sec:eom} and~\ref{sec:balance}, where we show that the energy lost by the mirror due to the optical back-reaction forces exactly matches the energy radiated into the field. Demonstrating this energy balance is the final objective of this paper. To this end, we move to the evaluation of the effective, semiclassical equations of motion for the mirror, in order to identify such optical forces.

\section{Mirror semi-classical equation of motion\label{sec:eom}}

\subsection{Optical back-reaction forces}

The influence action derived in the previous section allow us to construct the effective action governing the mirror dynamics, including the full back-action of the cavity field. Neglecting the first-order term $\mathcal{A}_1$, whose net contribution yields the static Casimir potential, which is well known and not the focus of this work, the effective CTP action takes the form:
\begin{multline}
    S_{\rm eff}[x_+,x_-] \;\equiv\; S_q[x_+] - S_q[x_-] + \mathcal{A}[x_+, x_-] \\= S_q[x_+] - S_q[x_-] + \mathcal{A}_2^I[x_+, x_-] + \mathcal{A}_2^{II}[\dot x_+, \dot x_-].\label{Seff}
\end{multline}
The equation of motion for the mean mirror displacement $x\equiv\langle\hat{x}(t)\rangle$ is obtained by varying the effective action with respect to either $x_+$ or $x_-$, and subsequently setting $x_\pm = x$:
\begin{equation}
    \frac{\delta S_{\rm eff}}{\delta x_+}\Big|_{x_\pm=x}  = \frac{\delta S_{\rm eff}}{\delta x_-}\Big|_{x_\pm=x} = 0.
\end{equation}
For definiteness, and without loss of generality, we perform variations with respect to $x_+$.
This procedure yields the equation of motion in the Euler–Lagrange form, which involves the effective Lagrangian $L_{\mathrm{eff}}$ corresponding to the effective action: $S_{\rm eff}=\int_{t_i}^{t_f}dt\,L_{\rm eff}$.
Note that only the following subset of the effective Lagrangian components contribute to variations with respect to $x_+$:
\begin{multline}
    L_{\rm eff}^+[x_+,x_-] \equiv L_q[x_+,\dot x_+] + \mathcal{L}_2^I[x_+, x_-] \\
    + \mathcal{L}_2^{II}[\dot x_+, \dot x_-].
\end{multline}
The equation of motion for the mean mirror displacement then follows from the Euler-Lagrange equation:
\begin{equation}
    \bigg(\frac{d}{dt}\frac{\partial L_{\rm eff}^+}{\partial \dot x_+} - \frac{\partial L_{\rm eff}^+}{\partial x_+} \bigg)_{x_\pm=x} = 0.
\end{equation}
Separating the free mirror dynamics from the field-induced back-action, this equation takes the instructive form
\begin{align}
    &\bigg(\frac{d}{dt}\frac{\partial L_q}{\partial \dot x_+} - \frac{\partial L_q}{\partial x_+} \bigg)_{x_\pm=x}  = F_x(t) + F_{\dot x}(t),\label{mirror_eom}
\end{align}
where we identified on the right-hand side the optical forces acting on the mirror due to the interaction with the cavity field. These are given explicitly as
\begin{equation}
    F_x(t)\;\equiv\;\frac{\partial\mathcal{L}_2^{I}}{\partial x(t)}\bigg|_{x_\pm={x}} = -\frac 4 d \int_{t_i}^t ds \;M_+^{00}(t-s)\, x(s),\label{Fx}
\end{equation}
and
\begin{multline}
    F_{\dot x}(t)\equiv-\frac{d}{dt}\frac{\partial\mathcal{L}_2^{II}}{\partial \dot x(t)}\bigg|_{x_\pm={x}} \\
    = \frac 1 d\int_{t_i}^t ds \; \Big(\dot M_{+}^{11}(t-s) \;+\; \dot M_{-}^{11}(t-s) \Big)\, \dot {x}(s).\label{Fx_dot}
\end{multline}
These forces are defined in terms of the $00$ and $11$ dissipation kernels and therefore share the same physical origin, namely the creation of excitation pairs in the field driven by the mirror motion. As discussed in Sec.~\ref{A_spectr}, this connection is most transparent in the frequency domain, which is the approach adopted in the following section, where the mechanical energy dissipated through these back-action forces is calculated.

\subsection{Mechanical energy dissipated by back-reaction forces}

As emphasized in the previous sections, a central objective of this work is to demonstrate that the CTP effective action formalism provides a self-consistent description of the nonequilibrium dynamics generated by the radiation–pressure interaction between the moving mirror and the optical field. A crucial test of this consistency is the explicit verification of energy conservation: The energy dissipated by the mirror through the nonlocal back-action forces must be exactly balanced by the energy gained by the field via pair-creation processes.
In this section, we derive the explicit expression for the mechanical energy loss. The energy gained by the field will be studied in Sec.~\ref{sec:balance}, where we switch to the in–out formalism in order to deduce this quantity from the total number of photons created in the field. The demonstrated agreement between these two calculations provides a definitive consistency check of the effective action approach.

The mechanical energy dissipated by the optical forces acting on the moving mirror is obtained by integrating the corresponding power:
\begin{align}
    E_{\rm diss}& = E_x + E_{\dot{x}} \nonumber\\
    &=\int_{t_i}^{t_f} ds\;\dot{x}(s)\,F_{x}(s) + \int_{t_i}^{t_f} ds\;\dot{x}(s)\,F_{\dot{x}}(s),
\end{align}
where $E_x$ and $E_{\dot{x}}$ are the contributions to the dissipated energy arising from the position- and velocity-dependent optical forces, respectively, and $\dot{x}$ is the mirror velocity, understood as the solution of the equation of motion, Eq.~\eqref{mirror_eom}. To decompose the dissipated energy into its frequency components, we take the limits $t_i \to -\infty$ and $t_f \to +\infty$ and work in Fourier space. Using Eqs.~\eqref{Fx} and~\eqref{Fx_dot}, together with the spectral decompositions of the dissipation kernels  $M_+^{00}$ and $M_\pm^{11}$ given respectively in Eqs.~\eqref{M00_spectr} and \eqref{M11_spectr}, we find that the energy dissipated by the position-dependent force $F_x$ receives contributions from the coupling of the mirror to individual field modes:
\begin{equation}
    E_x(\infty)=-\frac{2\pi^2}{d^2}\sum_{k=-\infty}^{+\infty} (2\omega_k)^3\,|\tilde x_{2\omega_k}|^2 {\rm Im}\,\tilde\mu_{kk}^{(2)},\label{E_x_discr}
\end{equation}
while the energy dissipated by the velocity-dependent force $F_{\dot{x}}$ receives contributions from the coupling of the mirror to pairs of distinct modes:
\begin{multline}
    E_{\dot{x}}(\infty)\\
    =-\frac{2\pi^2}{d^2} \sum_{\substack{k,j=-\infty \\ k\neq j}}^{+\infty}(\omega_k + \omega_j)^3 \,|\tilde x_{\omega_k+\omega_j}|^2\,{\rm Im}\,\tilde{\mu}_{kj}^{(2)}.\label{E_xdot_discr}
\end{multline}

By combining these results, the total mechanical energy dissipated by the back-reaction forces takes the form:
\begin{multline}
    E_{\rm diss}(\infty) =\\
    -\frac{2\pi^2}{d^2} \sum_{\substack{k,j=-\infty}}^{+\infty}(\omega_k + \omega_j)^3 \,|\tilde x_{\omega_k+\omega_j}|^2\,{\rm Im}\,\tilde{\mu}_{kj}^{(2)},
\end{multline}
with the equivalent expression obtained in the continuum limit:
\begin{multline}
    E_{\rm diss}(\infty) =
    -2\int_{-\infty}^{+\infty}d\omega \int_{-\infty}^{+\infty}d\omega' \,\\
    \times (\omega+\omega')^3 \, |\tilde{x}_{\omega+\omega'}|^2 \,{\rm Im}\tilde{\mu}_{\omega\omega'}^{(2)}.\label{E_diss_cont}
\end{multline}
Equation~\eqref{E_diss_cont} is a central result of this work. Its structure explicitly show that the mirror loses mechanical energy through the interaction with pairs of field modes. As stressed above, this admits a transparent interpretation in terms of the DCE, that is the spontaneous creation of real photon pairs from the quantum vacuum driven by the non-adiabatic motion of a boundary. Energy conservation in this process requires the mirror to supply the energy $\hbar(\omega+\omega')$ for each photon pair excited out of the vacuum, with frequencies $\omega$ and $\omega'$, respectively. This entails that the mirror loses energy precisely at the sum frequencies $\omega + \omega'$, in exact agreement with the factor $|\tilde{x}_{\omega+\omega'}|^2$ appearing in Eq.~\eqref{E_diss_cont}. This structure is the hallmark of a two-photon emission process and distinguishes the DCE from single-photon scattering, which would instead produce energy loss at frequencies $|\tilde{x}_\omega|^2$.

\section{Particles created in the field and energy balance\label{sec:balance}}

In the preceding sections, we studied the back-action of the cavity field on the mirror dynamics within the Feynman--Vernon influence functional framework~\cite{Feynman-IF}. This provided an open quantum systems description of the problem formulated in terms of the Keldysh--Schwinger CTP formalism. By tracing out the field degrees-of-freedom, we obtained an effective action for the mirror encoding back-reaction effects in the form of fluctuation and dissipation. From this effective action, we derived the semiclassical equation of motion for the mean mechanical displacement, explicitly identifying the optical back-action forces acting on the mirror. These forces are nonlocal, reflecting the causal structure of the underlying field correlations, and give rise to intrinsically non-Markovian dynamics. They introduce dissipation, and we calculated the mechanical energy lost by the mirror, interpreting this in the context of the DCE, as the energy radiated into the field through spontaneous photon pair creation.

A crucial consistency requirement of the effective action approach is energy conservation: the mechanical energy dissipated must exactly balances the energy transferred into the cavity field. While the CTP (in-in) formalism captures the field back-action, it does not directly access particle production, since the field degrees-of-freedom have been traced out and the framework is focused entirely on the mirror dynamics.

To establish the complementary field perspective, we turn now to the standard \emph{in--out} formalism. In this approach, particle production is captured by evaluating the mismatch between the initial (\emph{in}) and final (\emph{out}) states of the field driven by the mirror's motion. In what follows we show that, since the field is initially prepared in a thermal state at temperature $T$, this mismatch receives contributions from three distinct processes: spontaneous photon pair creation from the vacuum fluctuations, stimulated pair creation by the thermal photons already present in the cavity, and stimulated photon absorption by the mirror.

The net energy transferred into the field reflects the balance between these competing processes. While stimulated pair creation and stimulated absorption counteract and cancel one another, the net energy gained by the field is determined solely by spontaneous pair creation from vacuum fluctuations. This result highlights the purely quantum mechanical origin of the energy transfer. By computing the mean number of excitations created in each field mode with respect to the initial thermal state, we determine the total net energy transferred into the field and verify independently that it precisely matches the mechanical energy loss derived within the CTP approach. This agreement constitutes a nontrivial check of the consistency of the two formalisms and confirms that the DCE is the common physical mechanism underlying both the mirror dissipation and the field excitation.

\subsection{The \emph{in-out} formalism}

\subsubsection{Persistence amplitude of the initial state}

Due to its interaction with the moving mirror, the field evolves from the initial thermal state into a distinct final configuration that deviates significantly from equilibrium. As mentioned above, this mismatch results from the parametric amplification of both vacuum and thermal fluctuations, as well as the stimulated absorption of field excitations.

The degree of mismatch between the initial and final field states can be quantified through the so-called initial state persistence amplitude, which measures how much the state has persisted in its initial form. This quantity is most naturally introduced when the field is initially prepared in a pure state $|\Psi(t_i)\rangle$. In this case, the persistence amplitude is simply given as:
\begin{equation}
    A_{i\to f}[q] \equiv \langle\, \Psi(t_i) | \Psi(t_f) \rangle = \langle \Psi(t_i) | \hat{U}_{\mathrm{A}}^q(t_f,t_i)|\Psi(t_i) \rangle,\label{A_pers}
\end{equation}
and the state persistence probability is given by the squared modulus:
\begin{equation}
        P_{\rm pers} = A_{i\to f}[q]\, A_{f\to i}[q] = \big|A_{i\to f}[q]\big|^2.\label{pers_prob}
\end{equation}
If the final state coincides with the initial state, then $P_{\rm pers} = 1$, otherwise $P_{\rm pers} < 1$. This latter condition signals deviation from the initial equilibrium and therefore the occurrence of particle creation and/or annihilation in the field.

The generalization of Eq.~\eqref{A_pers} to the case of an initial mixed state is straightforward and reads:
\begin{equation}
    A_{i\to f}[q]= {\rm Tr}\big[\hat{\rho}_{\mathrm{A}}(t_i) \,\hat{U}_{\mathrm{A}}^q(t_f,t_i)\big].\label{A_pers_mix}
\end{equation}
Although, written in this form, the interpretation of this quantity as a persistence amplitude is less obvious and requires introducing the concept of thermofields~\cite{Takahashi1996,Barnett-thermofields-1985}, also known as quantum state purification in quantum information context~\cite{Barnett2009,Nielsen2010}. It is well known that it is possible to write any density operator $\hat{\rho}$ representing a mixed state in the Hilbert space $\mathcal{H}$, as a pure entangled state in the extended $\mathcal{H}\otimes \mathcal{H}$ space. This procedure effectively requires doubling the degrees-of-freedom of the problem by introducing an ancillary system $B$. Given the diagonal representation of the density operator, $\hat{\rho} = \sum_n\rho_n |\phi_n\rangle\langle \phi_n|$, a suitable purified state that reproduces the density operator defined above is the maximally entangled state:
\begin{equation}
    |\Psi_\rho\rangle = \sum_n \sqrt{\rho_n}\,| \phi_n\rangle \otimes | \phi_n\rangle_B.\label{pure_state}
\end{equation}
Here, $|\phi_n\rangle$ are the eigenstates of the density operator, while $|\phi_n\rangle_B$ identify the corresponding states in the ancillary Hilbert space. It is straightforward to demonstrate that original density matrix is recovered by tracing over the ancillary degrees-of-freedom.

In such an extended Hilbert space, Eq.~\eqref{A_pers_mix} takes the form
\begin{equation}
    A_{i\to f} [q]= \langle \Psi_\rho(t_i) | \hat{U}^q_{\mathrm{A}}(t_f,t_i) \otimes \mathbbm{1}_B |\Psi_\rho(t_i)\rangle,
\end{equation}
which is the persistence amplitude of the purified state $|\Psi_\rho\rangle$, with the evolution operator acting only on the physical field. This result demonstrates that Eq.~\eqref{A_pers_mix} can still be understood as a persistence amplitude of the initial state.

\subsubsection{In-out effective action and transition probability}

The probability for the field to persist in its initial state or transition to a distinct one is governed by the {in--out} effective action $\mathcal{W}[q]$, defined through the persistence amplitude as:
\begin{equation}
    A_{i\to f} [q]= e^{i\mathcal{W}[q]}.\label{A_W}
\end{equation}
Using this definition, the persistence probability reads:
\begin{equation}
    P_{\rm pers} = |A_{i\to f}|^2 = e^{-2{\rm Im}\mathcal{W}[q]},
\end{equation}
and the probability of transitioning to a distinct state is accordingly
\begin{equation}
    P_{\rm trans} = 1-P_{\rm pers} = 1- e^{-2{\rm Im}\mathcal{W}[q]}.
\end{equation}
If the radiation pressure interaction can be treated perturbatively, as is our working assumption, then the transition probability is small and can be approximated as:
\begin{equation}
    P_{\rm trans}\approx 2{\rm Im}\mathcal{W}[x].
\end{equation}
This result shows that the imaginary part of the in-out effective action directly encodes the total probability of exciting the field out of its initial state, and therefore describes particle production and absorption processes.

\subsection{In-out effective action for the radiation pressure interaction}

The relation between the {in--out} effective action $\mathcal{W}[x]$ and the {in--in} influence action $\mathcal{A}[x_+,x_-]$ follows directly from comparing their definitions in Eqs.~\eqref{A_pers_mix} and~\eqref{eq:IF}:
\begin{equation}
    \mathcal{W}[x] = \mathcal{A}[x_+=x,x_- = 0]. \label{W_vs_A}
\end{equation}
This relation admits a transparent physical interpretation. The functional $\mathcal{W}[x]$ describes the persistence amplitude of the initial field state and therefore involves only forward time evolution. In the Keldysh--Schwinger formulation, this corresponds to retaining solely the forward-propagating branch of the closed-time contour while setting the backward branch to zero.

Using the spectral representation of the influence action in Eq.~\eqref{A_FT}, the total probability of
exciting the field out of its initial state is:
\begin{align}
      &P_{\rm trans}\approx 2\,{\rm Im}  \mathcal{A}[x_+=x,x_- = 0]\nonumber\\[2pt]
      &\quad= 2\int_{-\infty}^{+\infty}d\omega \int_{-\infty}^{+\infty}d\omega'(\omega+\omega')^2\,\nu_{\omega\omega'}^{(2)}\,|\tilde x|^2_{\omega+\omega'},\nonumber\\[2pt]
      &\quad= 2\int_{-\infty}^{+\infty}d\omega \int_{-\infty}^{+\infty}d\omega'(\omega+\omega')^2\nonumber\\
      &\hspace{40mm}\times z_{\omega+\omega'}{\rm Im}\mu_{\omega\omega'}^{(2)}\,|\tilde x|^2_{\omega+\omega'}.\label{P_trans_rad_pres}
\end{align}
In the last equality, we used the fluctuation-dissipation relations $\nu_{\omega\omega'} = -iz_{\omega+\omega'}\mu_{\omega\omega'}$, together with the property of the dissipation kernel $\mu_{\omega\omega'}$ of being purely imaginary. 

Equation~\eqref{P_trans_rad_pres} quantifies the probability of a mismatch between the initial and final field states induced by the mirror motion. As discussed above, this mismatch arises from three distinct processes: spontaneous photon pair creation from the vacuum, stimulated pair emission, and stimulated absorption of thermal photons initially present in the field. All three contributions are encoded in the thermal factor $z_{\omega+\omega'} = 2n_{\omega+\omega'}(T) + 1$, where the term $2n_{\omega+\omega'}(T)$ accounts for stimulated emission and absorption, while the constant term $+1$ captures the vacuum contribution, i.e., spontaneous pair creation. Note that $z_{\omega+\omega'} \geq 1$ for all temperatures, reflecting the fact that thermal fluctuations always enhance the transition probability above its zero-temperature value.

Stimulated emission and absorption contribute equally and with opposite sign to the net energy transferred into the field, and therefore cancel exactly. The sole process responsible for a net energy transfer from the mirror to the field is spontaneous photon pair creation, that is the quantum vacuum contribution that survives even at zero temperature. To obtain the corresponding radiated energy, we weight  the integrand in Eq.~\eqref{P_trans_rad_pres} by the energy $\omega+\omega'$ (in units of $\hbar$) carried by each emitted photon pair, neglecting the stimulated contribution $ 2n_{\omega+\omega'}(T) $,  yielding the total mean energy injected into the field:
\begin{align}
      E_{\rm trans} &= 2\int_{-\infty}^{+\infty}d\omega \int_{-\infty}^{+\infty}d\omega'(\omega+\omega')^3|\tilde x_{\omega+\omega'}|^2{\rm Im}\mu_{\omega\omega'}^{(2)}\nonumber\\[2pt]
      &= -E_{\rm diss}.\label{E_trans}
\end{align}
As expected, this is equal and opposite to the energy dissipated by the optical back-reaction forces, Eq.~\eqref{E_diss_cont}. The equality $E_{\mathrm{trans}} = -E_{\mathrm{diss}}$ is the central result of this section. It establishes the exact balance between the mechanical energy lost by the mirror and the energy gained by the cavity field, thereby confirming energy conservation in our theory. This balance holds at the microscopic level of each individual created photon pair independently, demonstrating that each photon pair emission event is individually energy-conserving, as expected.

\section{Conclusions\label{sec:conclusions}}

In this paper, we have presented the in-in, or closed-time-path, functional formulation for a one-dimensional optical cavity bounded by two mirrors, one of which is free to move within a confining potential, enclosing a scalar field. The interaction between the field and the mirror has been modeled through Law's theory of the radiation pressure interaction. By calculating the Feynman--Vernon influence functional, we have derived an effective action for the mirror that encodes the full back-reaction of the field on the mirror dynamics, manifesting in the form of colored quantum noise and corresponding nonlocal dissipation. This non-Markovian dynamics is described by corresponding kernel functions in the influence action, which we have verified satisfy fluctuation-dissipation relations. We have analyzed the spectral structure of both kernels, revealing a direct connection with particle creation processes via the dynamical Casimir effect, thereby demonstrating that particle creation in the field is the physical mechanism responsible for the back-reaction. We have derived the semiclassical equations of motion for the moving mirror and identified explicit expressions for the optical back-reaction forces to which the mirror is subject. We have calculated the energy dissipated by these forces over the entire duration of the motion and demonstrated that it equals the energy transferred into the field by spontaneous particle creation. This explicit verification of energy conservation confirms the self-consistency of the formalism developed here.

As discussed in the Introduction, the formalism presented in this paper is intended as a first step toward a longer-term effort to develop theoretical tools for modeling the dynamics of complex, multi-mode condensed-matter and optical systems. While much of the existing work in these fields has relied on working assumptions that do not always reflect experimental reality, such as models characterized by a reduced number of optical modes~\cite{Savasta-PRX-2018,Savasta-PRL-2019,Butera-PRA-2019,Butera-EPL-2019}, or the use of semiclassical methods that fall short of capturing the dynamical evolution of quantum correlations~\cite{Butera-EPL-2019}, the approach we propose, and will continue to develop in future planned work, provides a comprehensive, first-principles description of the dynamics of optomechanical systems and, more generally, of quantum many-body systems~\cite{Rey-2004,Rey-2005,Rey-2006,Calzetta2007}.

\section{Acknowledgments}
The author thanks Bei-Lok Hu for being an endless source of inspiration, and Stephen Barnett and the members of the Quantum Theory Group at the University of Glasgow for their continuous support, stimulating discussions, and helpful exchanges.

\bibliography{TwoMirrors.bib}

\end{document}